\documentclass[letterpaper,twocolumn,10pt]{article}

\usepackage{usenix_2020_09}

\usepackage{algorithm}
\usepackage{algpseudocode}
\usepackage{amsmath} 
\allowdisplaybreaks
\usepackage{amsthm} 

\usepackage{amssymb} 
\usepackage{array} 
\usepackage{balance}
\usepackage{caption} 
\usepackage{comment}
\usepackage{enumitem}
\usepackage{graphicx} 
\usepackage{flushend}
\usepackage{hyperref}
\usepackage{listings}
\usepackage{makecell}
\usepackage{multirow}
\usepackage{pifont} 
\usepackage{siunitx}
\usepackage{stfloats} 
\usepackage{subcaption} 
\usepackage{tabularx}
\usepackage{textcomp}
\usepackage{tikz} 
\usetikzlibrary{positioning,fit}
\usepackage{wrapfig} 
\usepackage{xcolor}

\newcolumntype{C}[1]{>{\centering\arraybackslash}m{#1}}

\makeatletter
\algrenewcommand\alglinenumber[1]{%
  \makebox[1.5em][r]{\footnotesize #1:}\hspace*{0.25em}%
}
\makeatother

\newcommand{\cmark}{\ding{51}}
\definecolor{lightgray}{gray}{0.8} 

\definecolor{darkgray}{gray}{0.6} 

\newcommand{\xmark}{\ding{55}}

\definecolor{darkred}{RGB}{153, 0, 0} 

\newtoggle{our_comments}
\newtoggle{our_contents}
\toggletrue{our_comments}
\toggletrue{our_contents}

\newcommand{\akt}[1]{\iftoggle{our_contents}{{#1}}{}}

\newcommand{\RPCa}{RPC-A}
\newcommand{\RPCs}{RPC-S}
\newcommand{\RPCc}{RPC-C}
\newcommand{\abstraction}{RASC}
\newcommand{\abs}{RASC}
\newcommand{\sysname}{Rascal}
\newcommand{\dynpoller}{our dynamic polling strategy}
\newcommand{\scheduler}{DAG-TL}

\newcommand{\Figure}{Fig.}
\newcommand{\Table}{Table}
\newcommand{\Section}{Sec.}

\newtheorem{theorem}{Theorem}

\newcommand{\Figures}{Figs.}

\newtheoremstyle{thmparen}%
  {3pt}{3pt}
  {}{}
  {\bfseries}
  {.}
  {0.5em}
  {\thmname{#1}~\thmnumber{#2}\thmnote{ (#3)}}

\theoremstyle{thmparen}

\newtheorem{innercustomgeneric}{\customgenericname}
\providecommand{\customgenericname}{}

\newcommand{\newcustomtheorem}[2]{%
  \newenvironment{#1}[2][]
  {%
    \renewcommand\customgenericname{#2}%
    \renewcommand\theinnercustomgeneric{##2}%
    \innercustomgeneric[##1]%
  }{\endinnercustomgeneric}%
}

\newcustomtheorem{reptheorem}{Theorem}

\newcommand{\squishlist}
{
    \begin{list}{$\bullet$}
    {
        \setlength{\itemsep}{0pt}      \setlength{\parsep}{3pt}
        \setlength{\topsep}{3pt}       \setlength{\partopsep}{0pt}
        \setlength{\leftmargin}{1em} \setlength{\labelwidth}{1em}
        \setlength{\labelsep}{0.5em}
    }
}
\newcommand{\squishend}
{
    \end{list}
}

\newcommand{\squishenum}
{
    \begin{list}{\arabic{enumi}.}
    {
        \usecounter{enumi}
        \setlength{\itemsep}{0pt}      \setlength{\parsep}{3pt}
        \setlength{\topsep}{3pt}       \setlength{\partopsep}{0pt}
        \setlength{\leftmargin}{1.5em} \setlength{\labelwidth}{1em}
        \setlength{\labelsep}{0.5em}
    }
}

\lstdefinelanguage{yaml}{
  morekeywords={true,false,null,y,n},
  sensitive=false,
  morecomment=[l]{\#},
  morestring=[b]",
  morestring=[b]'
}

\lstdefinestyle{codeblock}{
  basicstyle=\ttfamily\scriptsize,
  numbers=left,
  numbersep=4pt,
  xleftmargin=9pt,
  breaklines=true,
  frame=lines,
  showstringspaces=false,
  columns=fullflexible
}

\begin{document}

\title{\abstraction{}: Enhancing observability \& programmability\\in smart spaces}

\author{
{\rm Anna Karanika, Kai-Siang Wang, Han-Ting Liang, Shalni Sundram, Indranil Gupta}\\
{\tt\{annak8,kw37,htliang2,shalnis2,indy\}@illinois.edu}\\
University of Illinois Urbana-Champaign\\
} 

\maketitle

\begin{abstract}

While RPCs form the bedrock of systems stacks, we posit that IoT device collections in smart spaces like homes, warehouses, and office buildings---which are all ``user-facing''---
require a more expressive abstraction.
{Orthogonal to prior work, which improved the {reliability} of IoT communication, our work focuses on improving the {\it observability} and {\it programmability} of IoT actions.}
We present the \abstraction{} (Request-Acknowledge-Start-Complete) abstraction, which provides acknowledgments at critical points after an IoT device action is initiated. \abstraction{}   is a better fit for 
IoT actions, which naturally vary in length {\it spatially} (across devices) and {\it temporally} (across time, for a given device).
\abs{} also enables the design of several new features: predicting action completion times accurately, detecting failures of actions faster, allowing fine-grained dependencies in programming, and scheduling.
\abs{} is intended to be implemented atop today's available RPC mechanisms, rather than as a replacement. We integrated \abs{} into a popular and open-source IoT framework called Home Assistant.
Our trace-driven evaluation finds that \abstraction{} meets latency SLOs,
especially for long actions that last  O(mins), which are common in smart spaces.
Our scheduling policies for home automations (e.g., routines) outperform state-of-the-art counterparts by $10\%$-$55\%$.
\end{abstract}

\thispagestyle{plain}
\pagestyle{plain}

\section{Introduction}
\label{sec:intro}

In the last 10 years, industry and users have moved from managing individual IoT (Internet of Things) devices to programming and managing {\it collections} of  IoT devices. 
IoT permeates our homes~\cite{IoTfuture}, workplaces~\cite{commit.mmsys},  farms~\cite{farmbeats}, Industry 4.0~\cite{lampropoulos2019internet}, entertainment venues~\cite{turchet2018internet}, etc. Smart buildings contain 2 B devices, expected to reach 2.5 B devices and a \$90 B market by 2027, and 4.12 B devices by 2030~\cite{Memoori,memoori2025iot}.
In a single deployment, 10s to 100s of IoT devices may be programmed and managed via automation containing a combination of actions~\cite{GHautomations,HAintegrations},
routines~\cite{ifttt,NodeRED},
IFTTT (If This Then That)~\cite{ifttt,reddit_ifttt_2021}, scripts, etc., 
~\cite{parks2025smarthome}.
An {\it action} is a single command sent to a single device, from a home hub (e.g., Alexa, Google Home hub) or a user device (e.g., phone, tablet, etc.). For instance, the Google Home API~\cite{GHautomations} contains over 50 defined actions, e.g.,  {\tt{blinds.OpenClose, SetFanSpeedRelative}, \tt{Cook}, \tt{Dispense}}, etc. 
The most common program written by users is a 
{\it routine}: a sequence of several actions,
which can be triggered conditionally by time, sensors, 
or manually by the user.

\begin{figure}[t!]
    \centering
    \includegraphics[width=\linewidth]{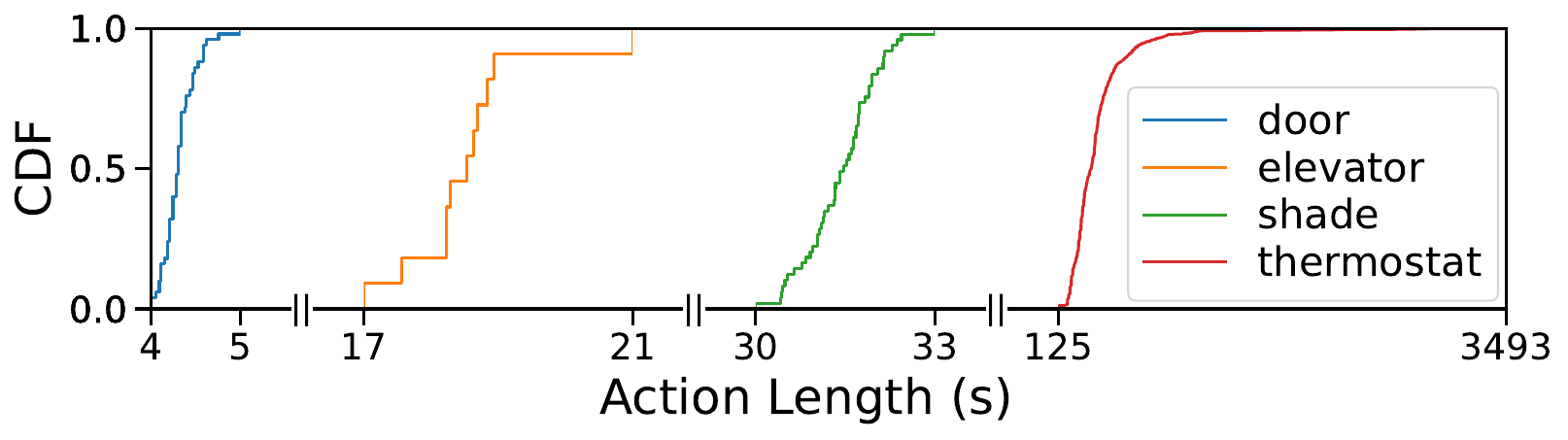}
    \vspace{-0.8cm}
    \caption{\bf \small IoT actions vary in length. {\it Actions: door: close, elevator: up 1 floor, shade: open, thermostat: heat  68.x to 69.y \textdegree Fahrenheit.}
    }
    \label{fig:action-length-variability}
    \vspace{-0.5cm}
\end{figure}

\noindent \textbf{RPCs and IoT Actions: }
Today, IoT actions
predominantly rely on the traditional RPC (Remote Procedure Call) abstraction~\cite{birrell1984implementing}.
RPCs are simple, widely understood, and accepted, and have many stock implementations~\cite{grpc,grpc_website,thrift2007,apache_thrift}.
An RPC consists of a single {\it Request} from a hub or
{personal device} towards the IoT device (or its proxy web service), and a single {\it Reply} in the reverse direction.

However, for IoT settings, it is challenging to map the Reply to the appropriate point in the action's execution. 
This is because in user-facing settings like smart spaces, many actions are non-instantaneous, taking seconds or even minutes to execute. 
Physical devices like windows and doors, shades and awnings, kitchen appliances, printers, etc., all 
take several seconds to execute an action (e.g., Google Home API's actions {\tt blinds.OpenClose, SetFanSpeedRelative}). That is, many actions in IoT settings are \textbf{\textit{long actions}}. 
Devices like sprinklers, ovens, HVAC systems (heating, ventilation, and air conditioning), etc., may take minutes or hours (e.g., Google Home API's actions {\tt Cook, Dispense,} etc.). Even ``fast'' devices like locks, light bulbs, fans
take a fraction of a second to finish an action. Together, these create a
non-trivial time gap between starting and completing an action. 
\Figure~\ref{fig:action-length-variability} shows several measurements we did of devices in our office buildings. 

Further, there is also a gap between the command's reception and start, due to
\textbf{\textit{preconditions}} for starting a command (e.g., coffee maker's reservoir
{must have} water). Finally, the \textbf{\textit{action may fail}} (e.g., main door lock is suddenly jammed) and create safety-violating situations. 

Returning to RPCs, 
this diversity in the nature and length of actions in IoT spaces raises the question of {\it which} version of RPC to use. Specifically---{\it when} is the Reply sent back? 
Today's RPC implementations are {\it forced to choose}  mapping of the Reply to one of three choices
---either: (1) {\it Acknowledgment}: sent back by the device or its
{cloud} service immediately when it receives the action (we call this {\it \RPCa}), but before the device has started executing the action, {\it or} (2) {\it Start} of the action: sent back by the device or its service when the action starts (we call this {\it \RPCs}), {\it or} (3) {\it Completion} of the action: sent back by the device or its service when the action is completed by the IoT device (we call this {\it \RPCc}). Many commercial smart home deployments today use  \RPCa{} combined with state updates~\cite{Alexa, GoogleHome}, while \RPCc{} is commonly employed in robotics settings~\cite{RECOVER_Cornelio, Zhang_2019_CVPR}.

\begin{table}
    \centering
    \footnotesize
    \setlength{\tabcolsep}{1pt} 
    \begin{tabular}{|c|p{3.5cm}||c|c|c|c|}
        \hline
        \textbf{Property} & \textbf{Goal} & \textbf{\RPCa} & \textbf{\RPCs} & \textbf{\RPCc} & \textbf{\abstraction} \\ \hline \hline
        {\it Observa}- & Progress updates (\Section~\ref{sec:adaptive-polling-strategy}) & \xmark & \xmark & \xmark & \cmark \\ \cline{2-6}
        {\it bility} & Failure detection (\Section~\ref{sec:detect-after-U}) & \xmark & \xmark & $\sim$ & \cmark \\ \hline
        {\it Program-} & Diverse dependencies (\Section~\ref{sec:causality}) & \xmark & \xmark & \xmark & \cmark \\ \cline{2-6}
        {\it mability} & Dynamic scheduling (\Section~\ref{sec:dyn-sched}) & \xmark & \xmark & $\sim$   & \cmark \\ \hline
    \end{tabular}
    \vspace{-0.3cm} 
    \captionof{table}
    {\bf \small Desired Properties. {\it  $\sim$ means non-trivial design needed.}
    }
    \label{tab:properties}
    \vspace{-0.4cm} 
\end{table}

\begin{figure}[t]
    \centering
        \begin{subfigure}[b]{\linewidth}
        \centering
        \includegraphics[width=0.55\textwidth]{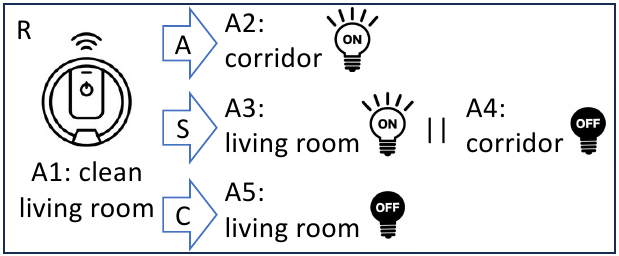}
        \caption{\small Single routine for energy-saving vacuum atop RASC.}
        \label{fig:energy-saving-vacuum-rasc}
    \end{subfigure}
    \begin{subfigure}[b]{\linewidth}
        \centering
        \includegraphics[width=\textwidth]{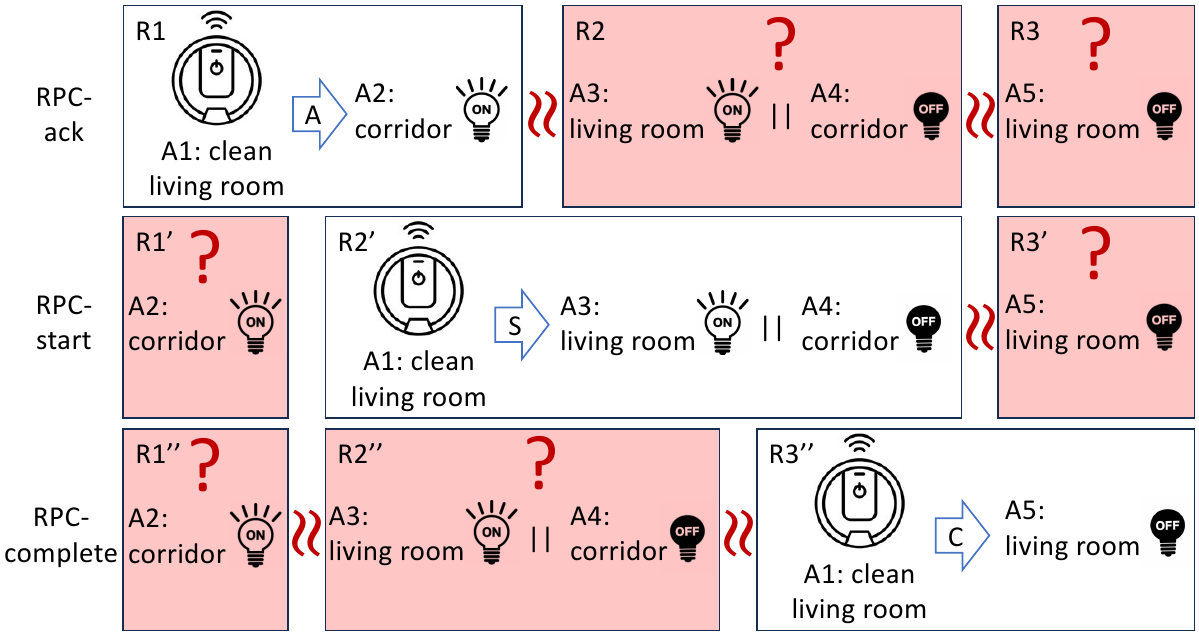}
        \caption{\small
        Ineffective RPC-based energy-saving vacuum routines. Pink routines need complex---or unavailable---triggers (\textcolor{darkred}{?}; e.g., ``vacuum in living room'' $\land$ ``cleaning/returning''). With suboptimal triggers, cross-routine conflicts (\rotatebox{90}{\bf \textcolor{darkred}{$\approx$}}) arise or actions fire unintentionally.
        }
        \label{fig:energy-saving-vacuum-rpc}
    \end{subfigure}
    \vspace{-0.7cm}
    \caption{\bf \small Routine example with dependencies on \textit{Acknowledgment} (A), \textit{Start} (S) and \textit{Completion} (C) among actions. {\it The vacuum requires light to navigate.}
    }
    \label{fig:energy-saving-vacuum}
    \vspace{-0.5cm}
\end{figure}

\noindent \textbf{Goals for an RPC Alternative: }
{The inherent {\it one-shot} nature of an RPC is a mismatch with the {\it long-running} nature of IoT actions.  We posit that the call-response for IoT device actions must adhere to two inter-related principles: 
\noindent (1) \textbf{\textit{Observability}}: the ability to track a device action.
\noindent (2) \textbf{\textit{Programmability}}: the ability to write and efficiently run {\it expressive} programs.

Observability implies that the hub (or user device) must receive \textbf{\textit{progress updates}} at critical points of the action execution. If any key part of the action fails, internal mechanisms must do  \textbf{\textit{fast failure detection}}, followed by mitigation. Second, programmability implies the ability for programs to support expressive and \textbf{\textit{diverse dependencies}} among actions (inside a program, and across programs), e.g., start the next action $A2$ earlier, {\it during} execution of a previous dependent action $A1$, after $A1$ has crossed key internal points (rather than waiting for $A1$ to finish). Diverse dependencies in turn require us to design \textbf{\textit{dynamic scheduling}} of the many actions that are executing in a smart space. Rescheduling is critical since action lengths have variance and sometimes a tail (\Figure~\ref{fig:action-length-variability}), and thus an action may take longer or shorter than its original estimation (and hence the original schedule). 
Table~\ref{tab:properties} summarizes these four goals.
This intertwined nature of observability and programmability also occurs in other domains like distributed systems~\cite{omnitable,viperprobe}. Kubernetes~\cite{kubern8observability}, and  SDN~\cite{simon}.

}

\noindent \textbf{A New Abstraction:}
This paper proposes {\it \abs} (Request-Ack-Start-Complete), a new abstraction for IoT devices, as an alternative to RPC. RASC does not replace RPC, but instead can be built atop RPC.
\abs{} provides Replies at three critical points of the action lifecycle
: {\bf A}ck (when the action is received by the device but not yet started), {\bf S}tart (when the device starts the action), and {\bf C}omplete (when the device finishes the action). Naturally, some of these replies may overlap   (e.g., if the action is short), but each is essential for longer actions
to satisfy \Table~\ref{tab:properties}'s goals. 

Consider a robotic vacuum that requires ambient light in its work area (for the robot's camera).
\Figure~\ref{fig:energy-saving-vacuum-rasc} shows a routine (a program with a set of actions) named $R$ expressed with diverse action dependency types. {Because the vacuum needs light for its camera to navigate, initially the corridor light is switched on so that the vacuum can find its way to the living room (since the precondition for the action is that the vacuum reach the living room). After that, the living room light is kept on for vacuuming, but the corridor light is switched off. When the vacuum has completed and docks in the living room, the living room light is switched off. } 
Similar routines exist involving lawn-mowers~\cite{santhoshini2024implementation}, tele-presence robots~\cite{tsui2011exploring}, etc.  

\Figure~\ref{fig:energy-saving-vacuum-rpc} shows that to {\it correctly} implement $R$ via RPC variants, $R$ needs to be split into several routines. 
Each split is problematic because (i) there are no good triggers for the pink routines (e.g., vacuum location cannot be used in trigger clauses), and (ii) if users go for clauses provided by the APIs, which tend to be simple, they will not get the intended result.

Our system \sysname{} implements the \abstraction{} abstraction. 
Assuring \Table~\ref{tab:properties} entails several challenges. 
First, we have to build a {\bf progress/failure detector} atop \abs, that minimizes detection time while respecting device constraints.
For poll-only devices,
it must poll often enough to track progress and catch failures promptly without overwhelming devices.
Second,
atop \abs{} when we provide support for expressive routine dependencies, we must innovate {\bf new dynamic scheduling algorithms} that reduce end-to-end latency (from routine arrival to completion).
Third, a key principle in our design is backwards compatibility: to make \sysname{} immediately deployable atop today’s ecosystems, IoT devices and their vendor services cannot be modified. This forces us to work within the constraints of existing RPC interfaces.
This is challenging as (i) \abs{} is more expressive than RPCs, and (ii) RPCs may go either via the IoT cloud service, or directly to the device.

This paper makes the following contributions:

\squishlist
\item We present a new expressive RPC-enhancing abstraction called \abs, for IoT settings.
\item We build the \sysname{} system that implements \abs{} over existing RPC-based IoT APIs. Concretely, to satisfy action progress updates and failure detection, we propose new techniques for efficient polling of the device. We also propose new dynamic scheduling policies that work with diverse action dependencies.
\item We integrate \sysname{}
with the popular and open-source home automation system called Home Assistant~\cite{HomeAssistant}.
\item We measured action execution times for various devices in office buildings. 
We use these and sets of real routines to perform trace-driven evaluation. We find that (i) \sysname{} 
detects completion within 2-13 RPCs and 2s-16s over 90\% of the time, and (ii)
our routine scheduling policies outperform state-of-the-art by 10\%-55\%.

\squishend

\section{System Model} 
\label{sec:model}

In today's deployments, devices communicate in one of five ways~\cite{HAdevicemodels}:
{\it Cloud Pull, Cloud Push, Local Pull, Local Push, Assumed State}. Here, {\it Cloud} means via the cloud service (device vendor's cloud service), while {\it Local} means locally directly from the device via Wifi, Bluetooth, Zigbee, etc. {\it Push} means the device updates the cloud service whenever its local state changes, while {\it Pull} means the cloud service has to poll the device for any state changes. For very old devices that allow neither pull nor push, the cloud service has to assume a state based on the last action. While the push variant naturally leans into providing updates for start and complete (and also intermediate states along the way), it is mostly adopted by sensor devices. Pull variants are widely prevalent---a cloud service or home hub continuously polls the device to detect updates. In fact, during our experiments, we have observed that at least 30\% of sampled vendors supported by Home Assistant only offer pull options.

The execution of a single action on an IoT device has three components: (1) a {\it network component} (transmission of request and response messages among the mobile, hub, cloud, device), (2) a {\it contextual component} (virtual and physical) comprising preconditions (either physical or safety-related) needed for the device to execute the action, and finally (3) a {\it physical component} (device executing the action). \abs{} basically argues that these three stages are spatially and conceptually distinct.
{
We assume IoT devices satisfy a few  properties: 

\squishlist
\item {\it IoT devices and their vendor services cannot be modified.} 

\item \noindent\textit{Only one action executes on a device at a time. } 
{This is typical for today's devices, some of which maintain a queue (e.g., HP printers~\cite{HP_PrintJobStuck_Queue}) 
while others reject new requests when busy (e.g., Nuki locks~\cite{Nuki423Locked}).}

\item \noindent\textit{Sufficient polling rate.} To query the state of a pull-based device, we assume arbitrarily frequent pulls are allowed, e.g., Philips Hue bulbs allow polling every second via Hue Bridge~\cite{HAHueRateLimit}. 

\item \noindent {\it Failures/Speed: } IoT devices may be arbitrarily slow or fail to execute actions. The network may delay or drop packets, and IoT devices and their vendor services need not be synchronized. 
We assume the reliability of devices beyond our purview (cloud services, hub, phones, etc.).
\squishend

\section{Design Goals}
\label{sec:properties}

We discuss \Table~\ref{tab:properties}'s key goals in detail. 

\noindent \textbf{Action progress updates.} 
This goal implies that 
the primitive should provide feedback at key points of the action execution: acknowledgment, action start, and action end (and if the device supports it, during the action). Doing so may require polling, and this polling needs to be efficient. 

\noindent \textbf{Action failure detection.} When an action fails, the primitive should detect it quickly.
This property may be implemented atop RPC-complete with the use of a timeout after the expected action length is over; however, that is not trivial since there is no fixed length for each action.

\begin{figure}[h]
    \vspace{-0.3cm}
    \begin{subfigure}[t]{0.48\linewidth}
        \centering
        {\includegraphics[width=0.95\textwidth]{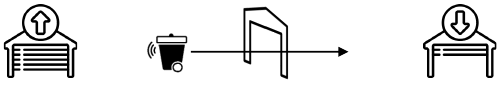}}
        \caption{\small \textit{Routine description:} (i) the garage door opens; (ii) the trashcan is moved out into the driveway; (iii) the garage door closes.}
        \label{fig:garage-routine-desc}
    \end{subfigure}
    \hspace{0.01cm}
    \begin{subfigure}[t]{0.47\linewidth}
        \centering
        {\includegraphics[width=\linewidth]{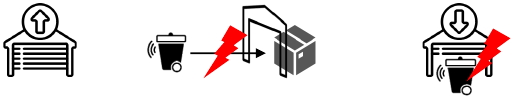}}
        \caption{\small \textit{Action failures:} the trashcan gets stuck below the garage door because of an obstacle and the garage door cannot close.}
        \label{fig:garage-routine-failure}
    \end{subfigure}
    \par\medskip
    \begin{subfigure}[t]{0.47\linewidth}
        \centering
        {\includegraphics[width=\textwidth]{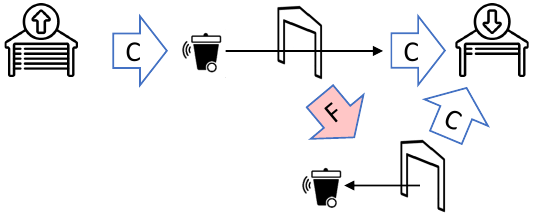}}
        \caption{\small \textit{Fallback actions:} the trashcan moves back inside after the first movement's failure and the garage door closes.}
        \label{fig:garage-routine-fallback}
    \end{subfigure}
    \hspace{0.01cm}
    \begin{subfigure}[t]{0.48\linewidth}
        \centering
        {\includegraphics[width=\textwidth]{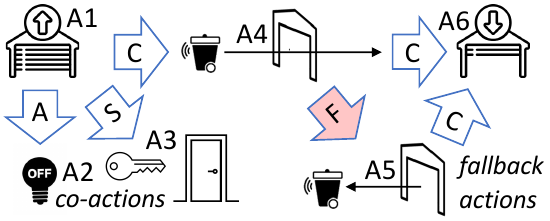}}
        \caption{\small \textit{Routine as a workflow:} diverse action dependencies allow for fine-grained control with \textit{co-actions} and \textit{fallback actions}.
        }
        \label{fig:garage-routine-workflow}
    \end{subfigure}
    \vspace{-0.3cm}
    \caption{\small \textbf{Routine with exception handling. {\it Garage door opens, automatic trash can~\cite{smartcan_protolabs}
    goes out, garage door closes. Letter inside arrow indicates trigger event: S=Start, C=Complete, F=Failure.}
    }}
    \label{fig:garage-routine}
    \vspace{-0.3cm}
\end{figure}

\noindent \textbf{Diverse action dependencies.}
The primitive must support rich inter-action dependencies. For a given example (garage door opens and closes to let the smart trashcan out), \Figures~\ref{fig:garage-routine}(a,b) show two scenarios, and \Figures~\ref{fig:garage-routine}(c,d) show two variants. 
In 
\Figure~\ref{fig:garage-routine-workflow},
actions $A2$ and $A3$ are \textit{co-actions} of action $A1$. The co-action concept allows (i) the light to turn off {\it before} the garage door starts opening, so that insects are not attracted inside, and (ii) the inside door to be locked only if the garage door starts opening; in case the garage door cannot open ($A1$ fails), there is no safety gap. 
If the trashcan move \textit{fails} en route to the driveway (\Figure~\ref{fig:garage-routine-failure}), the \textit{fallback action} $A5$ will move the trashcan to its original spot) and, finally, $A6$ will close the garage door 
to prevent exposure to robbers. 
If one were to implement \Figure~\ref{fig:garage-routine-workflow} via RPC, 
users must stitch together multiple routines (here, roughly four--one extra for every unsupported dependency type) guarded by complex device-state predicates to emulate progress points, and add extra conditions to block unintended triggers. This is prohibitive. 

\noindent \textbf{Dynamic action scheduling.}
When action lengths differ from the schedule, the primitive should reschedule the sequence of planned actions. 
Previous smart home scheduling for routines~\cite{SafeHomeEuroSys2021}  
(like \Figure~\ref{fig:garage-routine-desc}) ignores
unbounded  action lengths and diverse action dependencies (\Figures~\ref{fig:energy-saving-vacuum-rasc}, \ref{fig:garage-routine-workflow}), instead sticking to a static schedule.
Action durations are context-dependent (e.g., a vacuum may clean longer based on the floor area, dirt level, etc.), and failures are unpredictable at compile or schedule time. 
The schedule must change, ensuring: (a) \textit{safety}, i.e., no two actions trigger for the same device at the same time (which might cause the device to reject an action),  (b) action dependencies are upheld, and (c) if a previous action finishes early, the next dependent action can start early.
\RPCa{} and \RPCs{} cannot handle this, and \RPCc{} only allows us to detect when an action completes. \abstraction{} is more powerful. 
\section{Observability in \abstraction{}}
\label{sec:rasc}

The goals of action progress updates and failure detection (\Table~\ref{tab:properties}) can be served via a mechanism that {\it periodically polls} the device for its current state. This polling needs to (i) satisfy user-specified tolerances for detecting failures, (ii) balance detection latency vs.  device/network load, and (iii) adapt to evolving action time distributions (i.e., wrong estimates).

\begin{figure}[th]
    \centering
     \vspace{-0.3cm}
    \includegraphics[width=\linewidth]{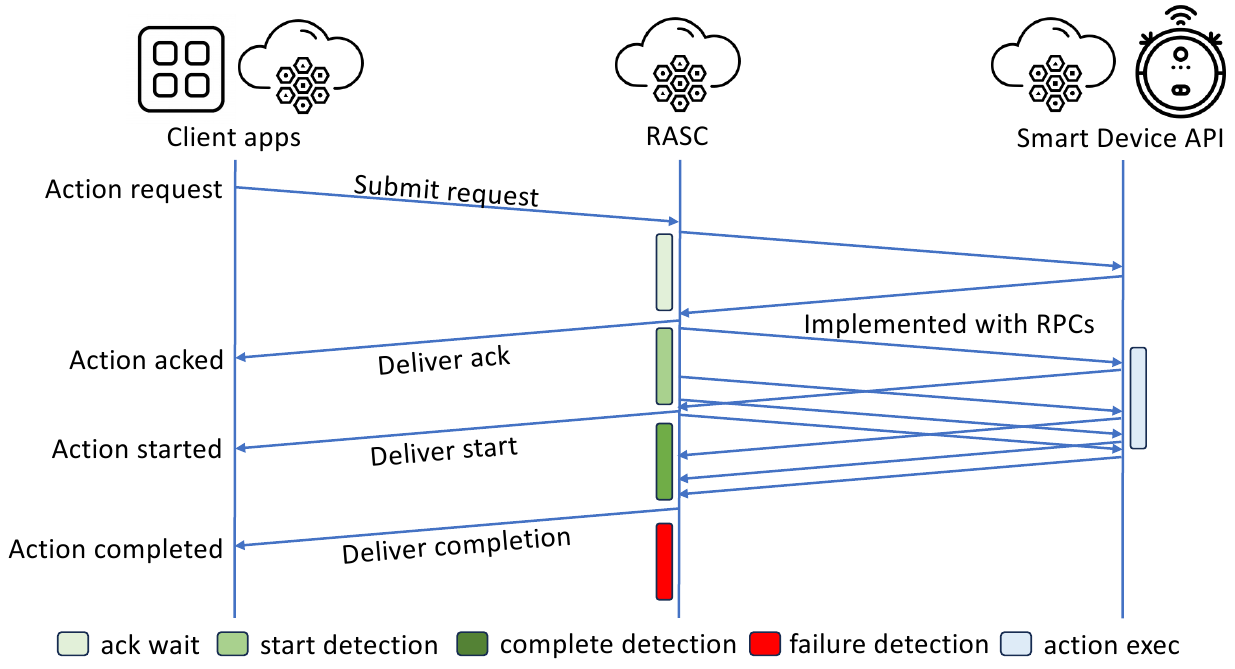}
    \vspace{-0.7cm}
    \caption{\bf \small \abstraction{} over RPCs. {\it Frequent RPCs on right are polls.}
    }
    \label{fig:rasc_design}
    \vspace{-0.3cm}
\end{figure}

\noindent \textbf{Lifetime of an Action.} \Figure~\ref{fig:rasc_design} shows how 
our \abs{} abstraction runs in a backward-compatible way over the RPC layer.  
For a given action, our  \sysname{} system can be in one of four states (or stages or phases, which we use interchangeably): {
{\it ack wait, start detection, complete detection}, and {\it failure detection}. 
When the action is requested, \sysname{} is in the {\it ack wait} state. Immediately after the hub receives an ack (that the device or its web service has received the request), \sysname{} enters the {\it start detection} phase, wherein it starts tracking the state of the device (described in \Section~\ref{sec:adaptive-polling-strategy}). 
Alternately, if the IoT device is unresponsive, then \sysname{} enters the {\it failure detection state}.  
The user specifies a {\bf tolerance threshold} $Q_w$ seconds (maximum time between failure and its detection), and \sysname{} needs to meet it. 

If the action is short, 
the action completion target state might be detected alongside or soon after the start.
So during the start detection state, \sysname{} also checks if either of the start or completion states is already matched, and if so, short-circuits to the complete detection state.

In IoT settings, it is impossible to distinguish a failed device from one that is not executing actions at all. So \sysname{} detects the failure of {\it actions}, but does not declare a device as being failed. \sysname{} reports failures to the user (who may take subsequent actions like restarting the device).

\begin{figure}[h]
    \vspace{-0.2cm}
    \centering
    \includegraphics[width=0.95\linewidth]{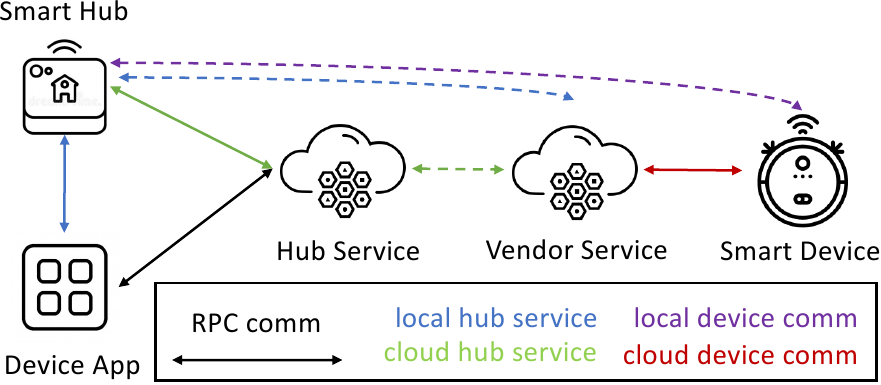}
    \caption{\bf \small Communication between a smart device and a client app. {\it Each arrowed line is an RPC in today's stacks. This paper replaces dashed lines with the new \abs{} abstraction.}
    }
    \label{fig:iot setup}
    \vspace{-0.3cm}
\end{figure}

\noindent \textbf{IoT Architecture.}
\Figure~\ref{fig:iot setup} shows how  \abs{} is built atop today's RPC mechanisms. 
\abs{} sits at the interface between user devices and applications on the one side (running on the Hub Service, e.g., Alexa, Google Home), and on the other side either the device or its vendor service  (e.g., web microservices run by the device vendor, e.g., Philips Hue, GE Sync, etc.).
In \Figure~\ref{fig:iot setup}, \abs{} replaces only those RPCs at dashed arrows, while solid arrow RPCs remain.
Every path from the hub/user device to the IoT device contains exactly one \abs{} call. 
{\abs{} needs to modify only the cloud services and hubs but not the device or vendor services.} 
\subsection{Adaptive Polling Strategy}
\label{sec:adaptive-polling-strategy}

Both progress updates and failure detection require \emph{polling} that is (i) efficient (few messages) and (ii) responsive (detects changes quickly).
For \emph{push} devices (\Section~\ref{sec:model}), this is straightforward---the device or vendor notifies the hub, which relays to \sysname{}. 
The challenge is \emph{pull}-only devices.

Polling too aggressively (e.g., every 500\,ms) can overload devices and the network; polling too sparsely increases detection latency. 
Key questions arise: {\it What polling frequency is appropriate? Should polling become more frequent as it progresses?}

\noindent \textbf{\underline{Problem Statement}.}
We seek to minimize \emph{detection time}: the gap between the action state change on a device and the next poll from \sysname{}.
Inputs are: (i) the {\bf historical state-change time distribution} \(p(t)\) for a (device, action) pair---\texttt{dist.pdf(t)}, (ii) a {\bf state change time bound} $U$---\texttt{dist.ppf(0.99)}\footnote{Percent point function (inverse cdf); \(\texttt{ppf}(y)\) returns the value with cdf \(y\).} which may occasionally be exceeded, and (iii) a poll \textbf{budget} \(k\) representing how many total polls are allowed for a given action (representing allowable bandwidth).
The goal is to select \(k\) poll timepoints in \((0,U]\) that minimize expected detection time.

\noindent \textbf{Solution.}
Let the $k$ poll times be $L_i, i=1,...,k$. Then the expected detection time $Q$ can be formulated as:

\vspace{-0.4cm}
\begin{align}
Q &= \int_{0}^{L_1} (L_1-t)p(t)\,dt 
   + \int_{L_1}^{L_2} (L_2-t)p(t)\,dt + \cdots \notag \\ 
  &+ \int_{L_{k-1}}^{L_k} (L_k-t)p(t)\,dt \tag{1} \\
Q &= \sum_{i=1}^{k} L_i \int_{L_{i-1}}^{L_i} p(t)\, dt
   - \int_{0}^{L_k} t\,p(t)\, dt \tag{2} \label{eq:Q}
\end{align}

Here, each term $i$ represents the expected detection time of $L_i$ if the change happens before $L_i$ and after $L_{i-1}$. The highest ${L_k}$ is always equal to $U$.
The second term here is the average of $p(t)$, and thus a constant we can ignore to optimize $Q$.

\begin{theorem}[Adaptive Poll Placement with Fixed Budget $k$]
Given a time distribution $p(t)$ on $(0,U]$, a poll budget $k$, and a terminal tolerance $\varepsilon>0$,
polls
$0<{L_1^{\!*}}<\cdots<L_{k-1}^{\!*}<L_k^{\!*}$
that \underline{minimize 
expected detection time} $Q$ and  satisfy $\lvert L_k^{\!*}-U\rvert\le\varepsilon$ are given by the following recurrent relation:
\[
L_i^{\!*} =
\begin{cases} 
\frac{1}{p(L_{i-1}^{\!*})} \cdot \int_{L_{i-2}^{\!*}}^{L_{i-1}^{\!*}} p(t)\, dt + L_{i-1}^{\!*} & \text{for } i \in \{2,\dots,k\}, \\
\text{value } \in (0,U] & \text{for } i=1
\end{cases} \tag{3} \label{eq:recurrence}
\]
\end{theorem}
\vspace{-0.5cm}

\begin{proof}[Proof Sketch (Full proof in Appendix~\ref{app:rasc-proofs})]
Based on \eqref{eq:Q}, only the $i$-th and $(i+1)$-th terms depend on $L_i$.
By applying a partial derivative to each $L_i$ and setting it equal to 0, we rearrange and get the recurrence relation \eqref{eq:recurrence}.
Under $p>0$ and continuity, these conditions determine a unique sequence once $L_1$ is fixed.
Next, we fix $L_1\in(0,U)$ and generate $L_2,\dots,L_k$ via \eqref{eq:recurrence}.
Then $L_k(L_1)$ is strictly increasing in $L_1$.
Hence, there exists a unique $L_1^*\in(0,U)$ such that the generated sequence satisfies $L_k^*=U$.
\end{proof}

\vspace{-0.2cm}

\begin{algorithm}[t!]
    \caption{\bf Adaptive Polling Algorithm}
    \label{algo:adaptive-polling-algorithm}
    \footnotesize
    \algtext*{EndIf} 
    \algtext*{EndElse} 
    \algtext*{EndFor} 
    \begin{algorithmic}[1]
        \Procedure{FindPolls}{dist, $U$, $Q_w$, $slo$}
            \State \Return RFindPolls(dist, $U$, 0, $ceil(U/Q_w)$,$Q_w$, $slo$)
        \EndProcedure
        \Procedure{RFindPolls}{dist, $U$, left, right, $Q_w$, $slo$}
            \State $k^*$ = (left + right) / 2
            \State $\mathcal{L}$ = GetPollingInterval(dist, $k^*$, $U$, left, right)

            \State valid = examineQw(dist, $\mathcal{L}$, $Q_w$, $slo$)

            \If{left == right $\And$ valid}
                \State \Return $\mathcal{L}$
            \EndIf

            \If{valid}
                \Comment{Try to reduce polls}
                \State \Return RFindPolls(dist, $U$, left, $k^*+1$, $Q_w$, $slo$)
            \EndIf

            \If{N+1 >= right}
                \Comment{Need more polls than $right$}
                \State \Return RFindPolls(dist, $U$, $k^*+1$, $right\times2$, $Q_w$, $slo$)
            \EndIf

            \State \Return RFindPolls(dist, $U$, $k^*+1$, right, $Q_w$, $slo$)
            
        \EndProcedure
        \Procedure{GetPollingInterval}{dist, $k$, $U$, left, right}
            \State $\mathcal{L}$ = zeros(k)
            \State $L_1$ = (left + right) / 2

            \State too\_large = False
            \For{i in range(2, $k$ + 1)}
                \State $L_i$ = calculateByRecurrenceRelation()
                \If{$L_i$ > $U$}
                    \State too\_large = True
                    \State \textbf{break}
                \EndIf
            \EndFor
            \If{isclose($L_k$, U)}
                \State \Return $\mathcal{L}$
            \EndIf
            \If{too\_large}
                \State \Return GetPollingInterval(dist, $k$, $U$, left, $L_1$)
            \EndIf
            \State \Return GetPollingInterval(dist, $k$, $U$, $L_1$, right)
        \EndProcedure
    \end{algorithmic}
\end{algorithm}

Next, to find the best sequence $\{L_i^{\!*}\}$
we use {\it binary search}. This leverages our empirical observations that there is a linear relationship between $L_1$ and $L_k$.
Algorithm~\ref{algo:adaptive-polling-algorithm} initially sets the left and right boundaries to 0 and $U$, respectively. In line 18, $L_1$ is set to the midpoint of the current left and right values. Using the recurrence equation from \eqref{eq:recurrence}, lines 20-29 calculate subsequent values based on $L_1$. We then check if $L_k$ is close enough to $U$.\footnote{The closeness is measured via the
{terminal tolerance $\varepsilon$}. 
{In our implementation, we set $\varepsilon=\num{e-5}$ (default in numpy np.isclose()).} 
}
If it is, we return $\{L_i\}$.  If not, we adjust the left and right boundaries and recursively search for the correct $L_1$.

After we obtain the placement $\{L_i^{\!*}\}$, 
we can examine the second derivative of each $L_i$, {where $L_0=0$ is a constant}:
\[
\small
L_i'' =
\begin{cases} 
2 \cdot p(L_i) - (L_{i+1} - L_i) \cdot p'(L_i) & \text{for } i \in [1,k-1], \\
2 \cdot p(L_i) + L_i \cdot p'(L_i) & \text{for } i=k
\end{cases} \tag{4}
\]

\noindent If any of them is negative, indicating a maximum instead of a minimum, the placement is invalid {and a larger $k$ is required}.

{\Figure~\ref{fig:poll} shows a pictorial example of a PDF for shade up, and the resulting polls generated by \sysname’s Algorithm~\ref{algo:adaptive-polling-algorithm}.}

\noindent \textbf{Meeting Detection Tolerance.} 
{We need to meet the user-specified {\it detection tolerance} $Q_w$ (gap between failure and its detection). 
More critical actions (e.g., involving locks, fire alarms, exhaust fans) use smaller values (e.g., < 1 s) while less critical actions (e.g., adjusting light brightness, the position of window shades, ovens) can use larger values.
We account for this by ensuring that $\forall i: L_{i+1} - L_i \leq Q_w$ {and $L_{i+1} - L_i >$ minimum polling interval allowed by the existing API (naturally, a $Q_w < $ min polling interval is unsupportable).}

Given $Q_w$, we use binary search to determine the smallest $k$ that meets $Q_w$---see  lines 4-15 in Algorithm~\ref{algo:adaptive-polling-algorithm}. Line 8  examines the $\mathcal{L}=\{L_i\}$ found by $GetPollingInterval$ to make sure any interval between any two polls is smaller than $Q_w$. If $\mathcal{L}$ is valid and left = right, we return  $\mathcal{L}$. If 
left does not equal right, this means we might be able to find  $\mathcal{L}$ 
with a smaller $k^*$. The algorithm recursively updates $k^*$ until it finds a $k^*$ that respects $Q_w$ and also cannot be smaller. The algorithm terminates when reducing $k^{*}$ by 1 would violate  $Q_w$.

\begin{figure}[t]
    \centering
    \includegraphics[width=0.9\linewidth]{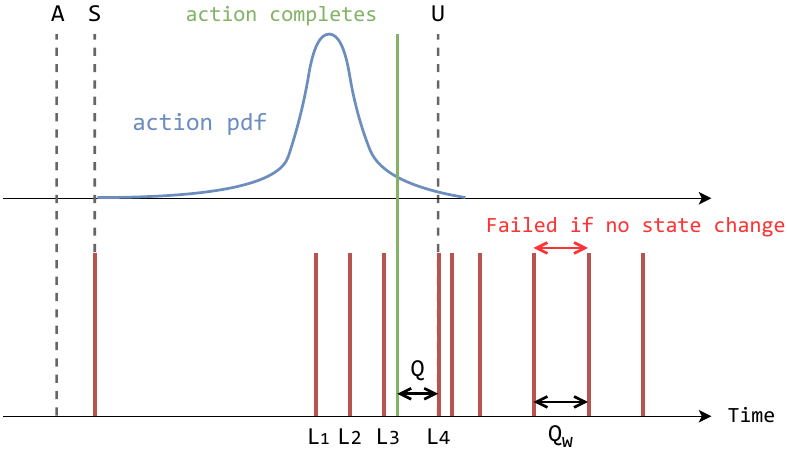}
    \vspace{-0.4cm}
    \caption{\bf \small Placement Example, $k = 4$. {\it Red lines are \abstraction{}  polls (after action start). Action completes between second and third poll.}}
    \vspace{-0.5cm}
    \label{fig:poll}
\end{figure}

\noindent \textbf{Reducing Polling further via  Confidence Intervals.} 
We relax the worst-case detection time assumption by introducing a {\it service-level objective (SLO)}: a confidence level for meeting the detection window $Q_w$. 
When SLO=0.9, \sysname’s placement may violate $Q_w$ on at most 10\% of events.
This allows \sysname{} to avoid excessive polling in high probability areas of the distribution.
Our evaluations (\Section~\ref{sec:detection_times}) find that  SLO-awareness does not change average detection time. Algorithm~\ref{algo:adaptive-polling-algorithm} shows how to find polls under SLOs.

\begin{theorem}[Meeting a Detection Tolerance SLO-Condensed version]
Given a detection window $Q_w$, a service-level objective $\text{SLO}\in(0,1]$, and a binary oracle that detects whether a $k$-poll placement with coverage of at least $\text{SLO}$ exists, binary search on the number of polls $k$ returns the smallest $k^*$ that meets the SLO. (\noindent Full theorem and proof in Appendix~\ref{app:rasc-proofs}.)
\end{theorem}
\vspace{-0.3cm}

\begin{proof}[Proof Sketch]
The oracle predicate is monotone: if some placement meets the SLO with $k$ polls, the SLO can still be met with any larger $k'$ by adding polls. The search is bracketed because $k=0$ is infeasible, while a sufficiently large $k$ (e.g., evenly spaced polls with $k=\lceil U/Q_w\rceil$) guarantees feasibility. Binary search on a monotone predicate with a valid bracket returns the minimal feasible $k^*$ 
that satisfies the SLO.
\end{proof}

\subsection{Detection Beyond  Upper Bound}
\label{sec:detect-after-U}

{
An estimate of the upper bound time for completing an action may be violated due to 
human interference (e.g., an elevator door is propped open), environmental factors (e.g., the heater takes longer when temperatures outside are colder), and out-of-order requests (e.g., the elevator receives a request to the 2nd floor when moving from 1st to 3rd). 
This means that
when no state change occurs by time $U$, polling must continue. 
The key question is: {\it 
What is the optimal post-$U$ polling technique?} 

Our key observation is that past $U$, the chance of an imminent state change typically declines, so \sysname{} polls densely at first and then tapers. 
Concretely, 
we distinguish between actions with progress (i.e., 
{{\it some} partial progress between critical points) and those without. For the former (e.g., vacuum has changed location), 
we estimate the remaining time from the instance's observed rate (time to the current progress) and thus schedule the next poll at $min\{\text{estimate},Q_w\}$.}
For the latter,
\sysname{} uses {\it exponential backoff}, doubling the inter-poll gap, capped by $Q_w$. 
This avoids overloading failing or troubled devices, and keeps worst-case detection  $< Q_w$ (\Figure~\ref{fig:poll}).
If no change occurs within $Q_w$ {after the upper bound $U$}
we declare the action failed. In our experiments, overruns beyond $U$ were rare, and late actions typically finished soon after $U$.
} 

\vspace*{-0.1cm} 

\section{Building Atop the \abstraction{} Abstraction}
While \Section~\ref{sec:rasc} addressed observability, we now turn to programmability: diverse dependencies and dynamic scheduling.

\begin{figure*}[t]
    \centering
    \hspace{0.1cm}
    \begin{subfigure}[b]{0.17\linewidth}
        \vspace{0pt}
        {\includegraphics[width=\textwidth]{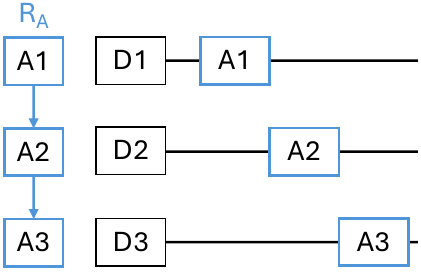}}
        \vspace{-0.5cm}
        \caption{\small $R_A$.}
        \label{fig:ra-sched}
    \end{subfigure}%
    \hfill
    \rule{1pt}{0.14\linewidth}
    \hfill
    \begin{subfigure}[b]{0.17\linewidth}
        {\includegraphics[width=\textwidth]{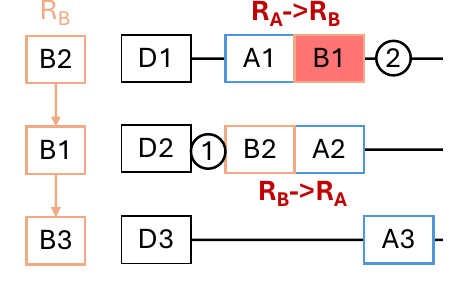}}
        \vspace{-0.6cm}
        \caption{\small $R_B$ \& invalid TL.}
        \label{fig:tl-invalid}
    \end{subfigure}%
    \hfill
    \rule{1pt}{0.14\linewidth}
    \hfill
    \begin{subfigure}[b]{0.19\linewidth}
        \vspace{0pt}
        {\includegraphics[width=\textwidth]{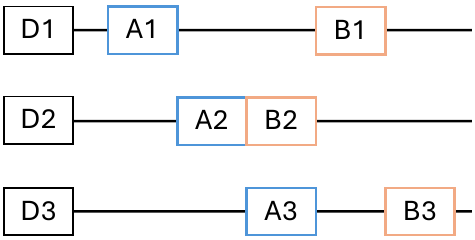}}
        \caption{\small $R_B$ \& valid TL.}
        \label{fig:tl-valid}
    \end{subfigure} 
    \hfill
    \rule{1pt}{0.14\linewidth}
    \hfill
    \begin{subfigure}[b]{0.2\linewidth}
        \vspace{0pt}
        {\includegraphics[width=\textwidth]{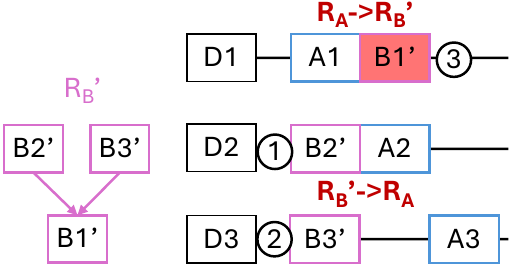}}
        \caption{\small $R_B'$ \& invalid \scheduler.}
        \label{fig:dag-tl-invalid}
    \end{subfigure}%
    \hfill
    \rule{1pt}{0.14\linewidth}
    \hfill
    \begin{subfigure}[b]{0.185\linewidth}
        \vspace{0pt}
        \centering
        {\includegraphics[width=\textwidth]{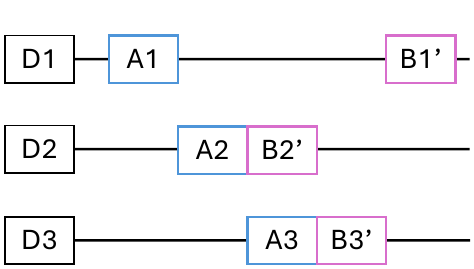}}
        \caption{\small $R_B'$ \& invalid \scheduler.}
        \label{fig:dag-tl-valid}
    \end{subfigure}
    \hspace{0.1cm}
    
    \vspace{-0.35cm}
    
    \caption{\bf \small Example Routines and Schedule for TL vs \scheduler. {\it Action symbols feature a letter that stands for the routine they are a part of and a number to denote which device they execute on, e.g., $A1$ is part of routine $R_A$ and executes on device $D1$.}}
    \label{fig:TL-vs-DAG-TL}
    \vspace{-0.45cm} 
\end{figure*}

\subsection{Diverse Action-Causality Expressiveness}
\label{sec:causality}

Commercial IoT platforms rarely support cross-action dependencies: e.g., Alexa routines allow limited inter-routine chaining,\footnote{Alexa can invoke one routine from another but offers no richer dependency model.} but not dependencies between actions. The \abs{} abstraction enables both cross-action and finer-grained dependencies across critical points {\it inside} an action's progress.
Figs.~\ref{fig:energy-saving-vacuum-rasc}, \ref{fig:garage-routine-workflow} depict {\it co-action} and \textit{fallback action} examples.

In \sysname{} each routine is represented internally as a DAG of device actions (\Figure~\ref{fig:energy-saving-vacuum-rasc}). When an action emits a progress event (A, S, or C), \sysname{} inspects its dependent children; any child whose dependencies are satisfied is immediately scheduled to run. \sysname{} execution is {\it event-driven}.

\subsection{Dynamic Action Scheduling}
\label{sec:dyn-sched}
We describe how to schedule routines and actions when some of these actions take a shorter or longer time than expected. 

\noindent\textbf{Background.}
In real-world IoT deployments, action durations vary widely:
ovens heat faster when pre-warmed, elevators stall if doors are obstructed,
and HVAC cycles fluctuate with outside temperature. 
Due to these variable times, any static schedule that is made, which devices execute (to respect dependencies) will need to be {\it dynamically} adjusted at runtime.

The real-time systems community studied this problem through
\emph{resource reclamation algorithms}~\cite{shen1993resource,manimaran1997new,gupta2000new}.
These algorithms optimize for \emph{early completions}:
when a task finishes ahead of its worst-case bound, subsequent tasks can
be advanced, reclaiming otherwise idle time without breaking precedence.
For example, the \emph{Restriction Vectors (RV)} approach~\cite{manimaran1997new}
maintains a timeline of tasks and, when a task finishes earlier than expected, pulls dependent tasks forward to begin once all of their parents have completed. This reclaims otherwise idle slack while preserving precedence constraints.

\noindent\textbf{Challenges.} 
We need to extend the reclaiming algorithms to not just account for actions that finish before their scheduled time, but also actions that take {\it longer} than their scheduled time.
A heater that lags on a cold day or a door that closes
more slowly than usual leaves later tasks delayed, and naive scheduling
can break causal dependencies or stall routines. 

\sysname's fine-grained causality model (\Section~\ref{sec:causality}) makes scheduling challenging:
we must honor dependencies at arbitrary progress points (e.g., Ack, Start, Complete), increasing potential conflicts and correctness risks.

Correct IoT scheduling hinges on two properties: \textbf{\textit{Safety}} (no device receives two action requests at once) and \textbf{\textit{Serial Equivalence}} (overlapping routines execute with some serial order). Without these, final states can be unpredictable or inconsistent with any one routine. SafeHome~\cite{SafeHomeEuroSys2021} guarantees these only when action durations are fixed.
However, with unpredictable durations and fine-grained dependencies, maintaining these guarantees requires adaptation.

\noindent\textbf{Overview.} 
\sysname{} extends reclamation-style scheduling to the IoT setting,
building atop the \abs{} abstraction (\Section~\ref{sec:rasc}).
\sysname's scheduling  consists of two phases:
(i)~it schedules arriving actions and routines (\Section~\ref{sec:scheduler}), and
(ii)~it dynamically adjusts the schedule to manage early action state change and delays (\Section~\ref{sec:rescheduler}). 
{
We present \textbf{\textit{DAG-TL}}, which assures serial equivalence for routines with variable action lengths. 
We then present two rescheduling policies, named \textbf{\textit{STF and RV}}, which 
adapt to both early and late state changes while preserving
Safety and Serial Equivalence.
}

\subsubsection{\scheduler: Assuring Routine Serial Equivalence}
\label{sec:scheduler}

We modify and adapt the Timeline Scheduler  (TL) from \cite{SafeHomeEuroSys2021}. 
When a new routine arrives, the original TL places actions in the earliest gaps that preserve the actions' order within a routine and the serialization order across routines.
TL uses an action-by-action backtracking approach to search through gaps that do not violate the established serialization order at any point in time. For \sysname-style DAGs, the original TL has exponential worst-case time complexity. 

{
To handle \sysname-style routines expressed as DAGs, our {\it \scheduler} approach 
adopts an {\it aggressive backtrack strategy} that makes ``big jumps'' in the state space.}
{Specifically, upon detecting a serialization conflict, it jumps to the routine’s top-level action and shifts the entire DAG forward in a single step---placing it after the last conflicting time found by inspecting the intersection of the preceding and following routine sets in the serialization order. By pruning unsatisfied paths early, we avoid repeated local retries.}

\noindent\textbf{Illustrative example.}
\Figure~\ref{fig:TL-vs-DAG-TL} shows three routines: $R_A$, $R_B$ and $R_B'$. After sequential $R_A$ is scheduled (\Figure~\ref{fig:ra-sched}), sequential $R_B$ arrives.
TL fills the earliest gaps: it tentatively places $B2$ (1), then places $B1$ (2), which reveals a serialization violation (\Figure~\ref{fig:tl-invalid}).
TL then moves $B2$ after $A2$ (the next available gap) and completes scheduling $R_B$ without further violations (\Figure~\ref{fig:tl-valid}).
Now consider non-sequential $R_B'$ arriving after $R_A$ (\Figure~\ref{fig:dag-tl-invalid}). 
Here $B1'$ depends on $B2'$ and $B3'$.
With a breadth-first traversal, \scheduler{} encounters a violation at the third step ($B1'$); it schedules $R_B'$ anew from the root and shifts the entire DAG so that all of $R_B'$ follows $R_A$ on any shared device (\Figure~\ref{fig:dag-tl-valid}). 
Naïve per-conflict backtracking would recheck and reschedule many prior actions and does not scale; \scheduler’s top-level backtrack reduces scheduling steps in practice.

\subsubsection{Rescheduler}
\label{sec:rescheduler}

An action may finish later (over-time) or earlier (under-time) than initially scheduled. \sysname{} must adjust.
We describe how actions are rescheduled to accommodate schedule deviations,
while minimizing routine completion time and continuing to ensure safety, serial equivalence, and performance.

{
\noindent \textbf{\underline{Problem Statement}. }
Each routine $R$ is a DAG of actions. An action $a$ has:
(i) a device $dev(a)$,
(ii) a (possibly data-driven) duration estimate $len(a)$, and
(iii) dependency edges to its parents in the routine $Pred_{\text{DAG}}(a)$.
Multiple routines may run concurrently and touch overlapping device sets.

Our goal is to maintain a {\it feasible schedule} that (1) is \textbf{Safe} (no device runs two actions at once), (2) is \textbf{Serially Equivalent}~\cite{SafeHomeEuroSys2021}  across routines (the final state equals some serial execution of whole routines), and (3) \textbf{adapts} when actions finish early/late while preserving (1)–(2).

}

\noindent \textbf{{When to Trigger the Rescheduler}. }
Late actions are handled by \sysname{} {\it proactively}, to assure safety. When a high percentage of the action's state change length upper bound $U$ (\Section~\ref{sec:adaptive-polling-strategy}) has elapsed (95\% in our implementation), \sysname{} extrapolates to estimate the new state change time (e.g., oven took 20 min to warm up from 200\textdegree F to 400\textdegree F, then \sysname{} estimates it will take 5 min more to get to the target 450\textdegree F).

Early-state-changing actions are handled
{\it reactively}, since early state change does not violate safety. If the difference between the expected and the actual state change time is above a threshold (1 s in implementation), we invoke the rescheduler. 

\noindent \textbf{{Constraints on the Rescheduler}. } 
The rescheduling has to respect two primary constraints: (1) dependency on prior actions inside the same routine, (2) and the (immutable) serialization order already established (\Section~\ref{sec:scheduler}) among active routines. We handle these via preprocessing and then two new rescheduling algorithms.

\noindent \textbf{{Preprocessing}.}
Let $\mathcal{I}$ be the set of potentially impacted actions. 
We add an action $A'$ to $\mathcal{I}$ when a deviation on action $A$ satisfies any one of three conditions:
(i) $A'$ is in the same routine $R$ and is a descendant of $A$ in $R$'s action DAG;
(ii) $A'$ is scheduled to start after $A$ on the same device $D$ that deviated;
(iii) $A'$ belongs to a routine $R'$ that is serialized after $R$.

To preserve the \emph{established} serial order among routines, \sysname{} records each routine’s \emph{postset}: the routines that, on any shared device, have executed after it up to the rescheduler’s trigger time. It then constructs an immutable serialization order by repeatedly: (1) selecting all routines with empty postsets, (2) placing them at the front of the order (breaking ties by arrival time), and (3) removing them from the remaining postsets. We iterate until all routines are placed.

\begin{figure}[h]
  \raggedleft

  \begin{subfigure}[b]{0.44\linewidth}
    \centering
    \includegraphics[width=\textwidth]{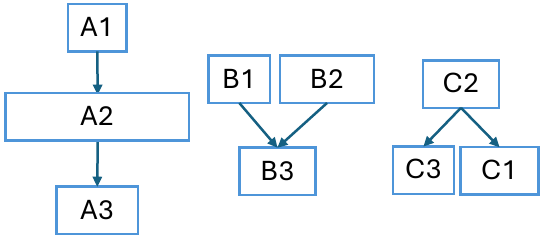}
    \caption{\small Routine graphs.}
    \label{fig:resched-routines}
  \end{subfigure}
  \hspace{0.1cm}
  \begin{subfigure}[b]{0.48\linewidth}
    \centering
    \includegraphics[width=\textwidth]{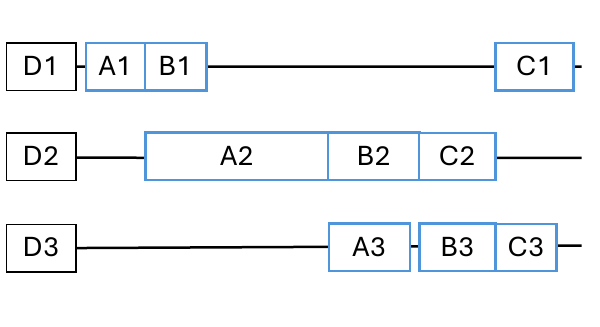}
    \vspace{-0.8cm}
    \caption{\small Original schedule.}
    \label{fig:resched-og-sched}
  \end{subfigure}

  \par\medskip

  \begin{subfigure}[b]{0.48\linewidth}
    \centering
    \includegraphics[width=\textwidth]{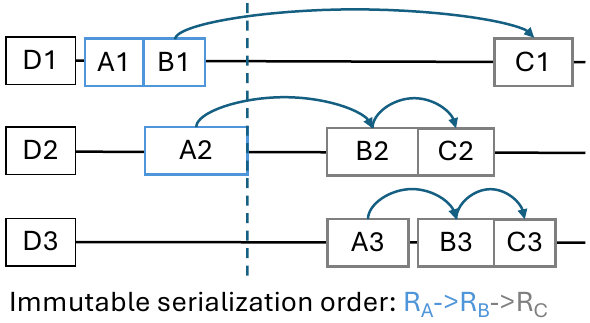}
    \caption{\small Preprocessing phase.}
    \label{fig:resched-a2-end}
  \end{subfigure}
  \hspace{0.1cm}
  \begin{subfigure}[b]{0.4\linewidth}
    \centering
    \includegraphics[width=\textwidth]{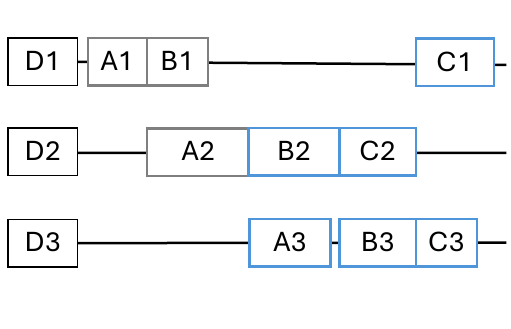}
    \caption{\small Rescheduling phase.}
    \label{fig:resched-sjfw-out}
  \end{subfigure}

  \caption{\small \textbf{Rescheduling in \sysname:} Example Routines.}
  \label{fig:resched}
\end{figure}

\sysname{} deschedules all actions in $\mathcal{I}$, and reschedules them while still preserving the established serial order.
For the routines in \Figure~\ref{fig:resched-routines} and the original \scheduler{} schedule in \Figure~\ref{fig:resched-og-sched}, \Figure~\ref{fig:resched-a2-end} shows the case where $A2$ completes early: affected actions (grey) are descheduled, the immutable order is $R_A\!\rightarrow\!R_B$, and routines that have not yet started (e.g., $R_C$) are left unchanged. Finally, \sysname{} inserts cross-routine, same-device dependencies to maintain serial equivalence. 

\begin{algorithm}[t]
    \caption{\bf Shortest Task First (STF) with Serializability}
    \label{algo:sjfw}
    \footnotesize
    \algtext*{EndIf} 
    \algtext*{EndElse} 
    \algtext*{EndFor} 
    \algtext*{EndWhile} 
    \begin{algorithmic}[1]
    \Require Impacted actions $\mathcal{A}$; duration $len(a)$; device $dev(a)$;
             routine DAG predecessors $Pred_{\text{DAG}}(a)$; immutable routine order $\prec$;
             device next-free times $next\_free[d]$ from current schedule.
    \Ensure Updated schedule without violating device safety or serial equivalence.
    
    \Statex \textbf{Phase 0: Add serialization edges (once).}
    \ForAll{pairs of routines $R \prec R'$}
      \ForAll{devices $d$ shared by $R$ and $R'$}
        \State Add edge from last action of $R$ on $d$ to first action of $R'$ on $d$
      \EndFor
    \EndFor
    
    \Statex \textbf{Phase 1: Initialize readiness and earliest feasible starts.}
    \ForAll{$a \in \mathcal{A}$}
      \State $Pred(a) \gets Pred_{\text{DAG}}(a) \cup Pred_{\text{serial}}(a)$
      \State $indeg[a] \gets |Pred(a)|$; \ $EST[a] \gets 0$; \ $FIN[a] \gets \text{unset}$
    \EndFor
    \State $Ready \gets \{ a \in \mathcal{A} \mid indeg[a]=0 \}$
    \ForAll{$a \in Ready$}
      \State $start\_cand[a] \gets next\_free[dev(a)]$
    \EndFor
    
    \Statex \textbf{Phase 2: List scheduling (single priority queue).}
    \While{$Ready \neq \emptyset$}
      \State $a^\star \gets \arg\min_{a \in Ready} \ (start\_cand[a], len(a))$ \Comment{Earliest start, then shortest task}
      \State $s \gets start\_cand[a^\star]$; \ $f \gets s + len(a^\star)$
      \State Place $a^\star$ on timeline of $dev(a^\star)$ at $[s,f)$
      \State $FIN[a^\star] \gets f$; \ $next\_free[dev(a^\star)] \gets f$
      \State Remove $a^\star$ from $Ready$
      \ForAll{successors $b$ of $a^\star$ in $Pred(\cdot)$}
        \State $indeg[b] \gets indeg[b]-1$
        \State $EST[b] \gets \max(EST[b], \ FIN[a^\star])$
        \If{$indeg[b]=0$}
          \State $start\_cand[b] \gets \max(EST[b], next\_free[dev(b)])$
          \State Insert $b$ into $Ready$
        \EndIf
      \EndFor
    \EndWhile
    \State \textbf{return} updated schedule
\end{algorithmic}
\end{algorithm}

\noindent \textbf{{Rescheduling}.} We reschedule the descheduled actions using two policies.
Our first policy is the Shortest Task First (STF) algorithm (known to be optimal~\cite{Tanenbaum2006}
). We augment it to honor dependencies within and across routines (Algorithm~\ref{algo:sjfw}).
keyed by estimated length; we repeatedly (i) pick the head, (ii) place it at the earliest common free time across its required devices, and (iii) update downstream dependencies, until the queue empties.
In \Figure~\ref{fig:resched}'s example, respecting the immutable serialization order, our STF variant does not place $C2$ before $B2$ even though $C2$ is shorter (\Figure~\ref{fig:resched-sjfw-out}).

\begin{figure}[t!]
    \vspace{-0.4cm}
    \begin{subfigure}[t]{0.18\linewidth}
        \centering
        \includegraphics[width=\linewidth]{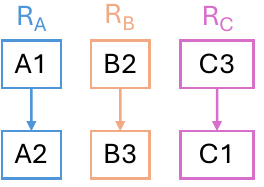}
        \caption{\small Routine graphs.}
        \label{fig:rv-graphs}
    \end{subfigure}
    \hfill
    \begin{subfigure}[t]{0.26\linewidth}
        \centering
        \includegraphics[width=\linewidth]{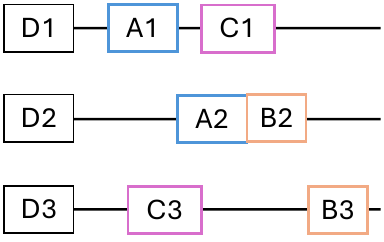}
        \caption{\small Original \scheduler{} schedule.}
        \label{fig:rv-dag-tl}
    \end{subfigure}
    \hfill
    \begin{subfigure}[t]{0.25\linewidth}
        \centering
        \includegraphics[width=\linewidth]{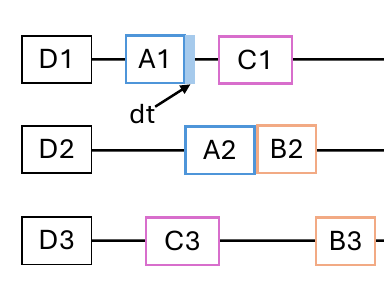}
        \caption{\small $A_1$ completes early $\rightarrow$ only RV.}
        \label{fig:rv-early}
    \end{subfigure}
    \hfill
    \begin{subfigure}[t]{0.28\linewidth}
        \centering
        \includegraphics[width=\linewidth]{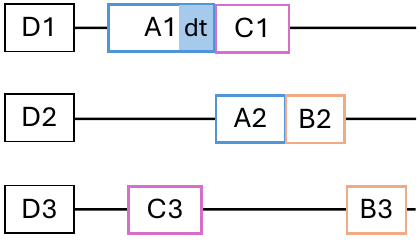}
        \caption{
            \small $A_1$ completes late $\rightarrow$ Push + RV.
        }
        \label{fig:rv-late}
    \end{subfigure}
    \vspace{-0.3cm}
    \caption{\bf \small RV rescheduling. {\it For (d), upon $A_1$'s late completion detection by an estimated $dt$, $C_1,A_2,B_2,B_3$ are first pushed forward by $dt$ before RV is performed.}}
    \label{fig:rv}
    \vspace{-0.5cm}
\end{figure}

{
Our second policy is based on \emph{Restriction Vectors (RV)} from the real-time systems  community~\cite{manimaran1997new}.  
The classical RV pulls a task forward as soon as all its parents finish to reclaim slack.
In \sysname{}, we apply this directly when progress points (e.g., \textit{Start}, \textit{Complete}) occur earlier than scheduled.
Unlike classical real-time settings, IoT actions may also finish \emph{later}, which is acceptable; to remain correct, we extend RV as follows:
(i) shift \textit{all} subsequent, not-yet-started actions forward by the observed delay to preserve safety; then
(ii) apply RV to the shifted schedule to exploit any remaining slack.
Thus, RV stays correct under late completions while still harvesting early-completion slack (see \Figure~\ref{fig:rv}).
}

{

\begin{theorem}[Action Safety]
\label{thm:safety}
{For any device $D$, \sysname{}: (i) never initiates $>1$  action to $D$ concurrently, and  (ii) initiates an action only if $D$ is idle. Thus, ack-ed actions on $D$ execute in isolation.
}
\end{theorem}
\vspace{-0.5cm}
\begin{proof}
\emph{\scheduler} assigns each action to the earliest idle slot on $D$ that respects dependencies.  
\emph{RV} may move actions earlier, but only when $D$ is idle and all parents are complete.
\emph{STF} inserts each action into the earliest idle slot consistent with serialization. Neither rule permits overlapping initiations.  
Thus, the non-overlap invariant is preserved under both schedulers and all rescheduling steps.
\end{proof}

\begin{theorem}[Serial Equivalence]
\label{thm:serial}
For any set of concurrent routines $\mathcal{R}$, the schedule $\mathcal{S}$ produced by \sysname{} is equivalent to executing $\mathcal{R}$ in some serial order $\pi(\mathcal{R})$.
This holds for the baseline scheduler and for rescheduling under RV and STF.
\end{theorem}
\vspace{-0.5cm}
\begin{proof}
\emph{\scheduler:} enforces a partial order $\prec$ on conflicting routines and produces a schedule $\mathcal{S}$ that extends $\prec$.
\emph{RV} will
advance or delay an action’s start $s(a)$ only if $\prec$ is preserved.
\emph{STF} places each action $a$ in the earliest idle slot $i(a)$ consistent with $\prec$. 
No step moves an action past the serialized prefix of a conflicting routine;
thus, $\mathcal{S}$ is equivalent to some serial order $\pi(\mathcal{R})$ consistent with $\prec$.
\end{proof}
}

\section{Implementation}

We implemented \sysname{} as a new component for Home Assistant (HA)~\cite{HAcore}, a widely used open-source smart-home platform. The prototype comprises \(\sim\)8.5K Python LOC. 
\akt{Deploying \sysname{} requires no changes to the HA core, existing integrations, or devices and their APIs. The user only needs to install \sysname{} and 
supply lightweight per-device-class progress mappings that bind each action's key execution milestones (i.e., start, completion) to observable fields the class already exposes (state attributes). This lets \sysname{} infer action progress and schedule polls without modifying devices.}
We also built a 500-LOC event-driven HA simulator and adapted the \texttt{hass-virtual} third-party component~\cite{hassvirtual} for device simulation and Raspberry Pi-based emulation.

\begin{figure}[t]
\centering
\begin{minipage}{\linewidth}
\begin{lstlisting}[style=codeblock,language=yaml]
- id: "1715802493830"
  alias: Heat when window starts closing and door is closed
  action:
    - parallel:
        - service: cover.close_cover
          target: { entity_id: cover.living_room_window }
        - service: cover.close_cover
          target: { entity_id: cover.balcony }
    - service: climate.set_temperature
      data: { temperature: 72 }
      depend_on: [start, complete]
      target: { entity_id: climate.main_thermostat }
\end{lstlisting}
\vspace{-0.5cm}
\caption{\bf \small Example RASC routine with \texttt{depend\_on} specification. 
{\it \texttt{desc}, \texttt{trigger},  \texttt{condition} omitted for brevity. The main thermostat is requested to heat to 72\textdegree F after the living room window has started closing \& the balcony door has closed.
}}
\vspace{-0.5cm}
\label{lst:script-example}
\end{minipage}
\end{figure}

\noindent \textbf{\abs{} API.}
We extend HA’s action YAML with a single optional field, \texttt{depend\_on}. 
This field is an \emph{ordered list} of required progress events---one per parent, in the parents’ order (e.g., \texttt{ack}, \texttt{start}, \texttt{complete})---enabling fine-grained sequencing.
An example routine with mixed dependencies appears in \Figure~\ref{lst:script-example}.
If \texttt{depend\_on} is omitted, we default to \texttt{complete}, preserving backward compatibility with HA’s sequential semantics.
At runtime, the YAML is compiled into a dependency DAG; \scheduler{} monitors this DAG and triggers actions as soon as prerequisites are satisfied.

\noindent \textbf{Action length PDF.} 
We obtain each action's probability distribution function by storing historical durations for (ack→start) and (start→complete).
Actions are keyed by \(\langle\textit{device},\,\textit{action},\,\textit{transition}\rangle\). 
Consequently, \dynpoller{} requires a brief training phase (several trials) before it can return accurate results.
{
\noindent \textbf{Asynchronous poll placement computation.}
Algorithm~\ref{algo:adaptive-polling-algorithm}'s compute time increases with the upper bound, which can be large for long actions. 
To avoid polling delays
in \sysname, we decouple computation from execution and make it asynchronous: upon an action's completion, poll placement is computed anew and stored for its next initiation. 
}

\noindent \textbf{Action status change notification.}
On each status change, \sysname{} publishes an \texttt{\abs\_RESPONSE} event on HA’s event bus; applications subscribed to this topic receive the update.

\vspace*{-0.3cm} 

\section{Experimental Evaluation}
\label{sec:eval}

We address the following research questions:
\squishenum
    \item How efficient is \sysname's polling technique? (\Section~\ref{sec:detection_times})
    \item Does action length estimation converge? (\Section~\ref{sec:training-eval}) 
    \item If the action length distribution evolves, can \sysname{} reconverge its learnt distribution quickly? (\Section~\ref{sec:training-eval})
    \item When actions are interrupted, can \sysname{} distinguish them from a scenario where they actually fail? (\Section~\ref{sec:interruption-eval})
    \item Under realistic routines, how do 
        \scheduler+\{RV, STF\} compare to baselines~\cite{SafeHomeEuroSys2021} {in latency \& overhead}? (\Section~\ref{sec:sched-finetuning})
\squishend

\noindent {\bf Trace Collection.} 
To drive our trace-driven emulation, we collected traces of (I) device action times that form a diverse set of devices in an office building, and (II) a set of diverse routines, inspired by existing routine datasets including IoTBench~\cite{iotbench,smartthingsapps}, described in \Section~\ref{sec:sched-finetuning}.
For (I), we manually collected a significant dataset of action completion times. We recorded multiple ($>50$) trials over 5 days, with 2 elevators, 2 projector screens, 2 lights, 2 doors, and 4 shades. Elevator data was collected while other users were using it, and thermostat traces are from an existing dataset~\cite{fraunhofer_thermostat_2022}.

\subsection{\sysname{} Polling Efficiency}
\label{sec:detection_times}

We compare \sysname's adaptive polling efficiency to a baseline periodic polling strategy, which polls every $Q_w$.
\Figure~\ref{fig:polls-per-action} shows that compared to periodic polling, adaptive polling uses 44\% to 89\% fewer polls to detect the action completion. 
We set the Service Level Objective or $SLO$---the percentage of events \sysname{} promises to detect within $Q_w$ (Section~\ref{algo:adaptive-polling-algorithm})---
to 0.9 and $Q_w$ equal to 2, 3, and 30 (seconds) for door, shade, and thermostat actions, respectively. 
Action \texttt{shade down}'s detection time is higher with adaptive polling but still below $Q_w$. This data point highlights the following tradeoff: periodic polling is costly and guarantees detection within $Q_w$ always, whereas adaptive polling is very cost-efficient but may detect action progress within $Q_w$ only with high probability ($\ge SLO$). 

\begin{figure}
    \centering
    \includegraphics[width=\linewidth]{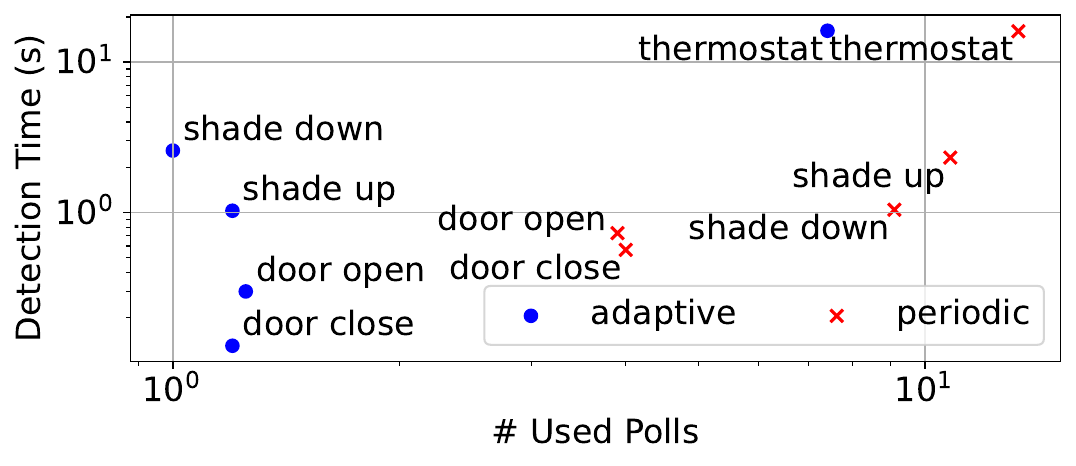}
    \vspace*{-0.8cm}
    \caption{\bf \small \sysname's Adaptive Polling vs. Baseline Periodic Polling. \it \sysname{} is better (lower and to the left) in 4 of 5 cases. 
    }
    \label{fig:polls-per-action}
    \vspace*{-0.3cm}
\end{figure}

\begin{table}[t]
    \centering
    \footnotesize
    \setlength{\tabcolsep}{2pt}
    \renewcommand{\arraystretch}{1.02}
    \begin{tabular}{|l||r|c|c|c|c|r|}
        \hline
        \multirow{2}{*}{\textbf{Action}} & \centering \textbf{Avg.} & \multicolumn{2}{c|}{\textbf{Detection Time}} & \multicolumn{2}{c|}{\textbf{Computation Time}} & \multirow{2}{*}{\textbf{Speedup}} \\
        \cline{3-6}  
           & \textbf{Length} &      Adaptive &  V-opt         & Adaptive          & V-opt     & \\
        \hline\hline
        Door close  &   3.19 & 0.19          & \textbf{0.13} & \textbf{6.04e-04} & 4.38e-02 & 59.37 \\ \hline
        Door open   &   3.06 & \textbf{0.19} &  0.30         & \textbf{6.43e-04} & 3.82e-02 & 72.56 \\ \hline
        Shade up    &  29.64 & \textbf{0.31} &  1.02         & \textbf{3.96e-02} & 1.97e+01 & 497.77 \\ \hline
        Shade down  &  27.45 & \textbf{1.54} &  2.58         & \textbf{2.93e-02} & 1.63e+01 & 554.99 \\ \hline
        Therm 68,69 & 432.17 &      --       & 16.19         & \textbf{9.04}     & {2hr+}   & 796.34 \\ \hline
    \end{tabular}
    \vspace{-0.3cm} 
    \caption{\bf \small {\sysname's Adaptive Polling vs. Baseline V-opt-calculated Polling, Detection vs. Computation Time. All times are in seconds. Bold font indicates the lowest value for each action/metric combination. \it \sysname{} has similar detection times, but V-opt's computation time is impractical, especially for long actions.}}
    \label{tab:V-opt}
    \vspace{-0.5cm} 
\end{table}

{
Table~\ref{tab:V-opt} compares \sysname's adaptive polling against the baseline called V-optimal (V-opt)~\cite{ioannidis1995balancing},
the classic dynamic-programming algorithm for histogram construction. V-opt selects bucket boundaries to minimize variance across bins.
Both achieve comparable detection times, but \sysname{} computes poll placements 60-800$\times$ faster than V-opt.  
This efficiency gap makes V-opt impractical in real deployments; its computation time can exceed even the average action length (e.g., thermostat), while \sysname{} completes in milliseconds.
}

\vspace*{-0.3cm} 

\subsection{\sysname{} Training}
\label{sec:training-eval}

\noindent \textbf{Convergence Speed.} 
We measure how many action-length samples \sysname{} needs to reach a {\it stable} per–(device, action) distribution.
We define ``stable'' as: mean and variance both differ by less than 5\% upon consecutive samples.
\Figure~\ref{fig:convergence} shows that 10-20 samples suffice across devices: actions lasting tens of seconds (door, elevator, shade, projector screen) converge in $\sim10$ samples, while longer actions (e.g., the 400\,s thermostat) require $\sim20$.

\begin{figure}[t!]
    \centering
    \begin{subfigure}{\linewidth}
        \centering
        \includegraphics[width=0.8\linewidth]{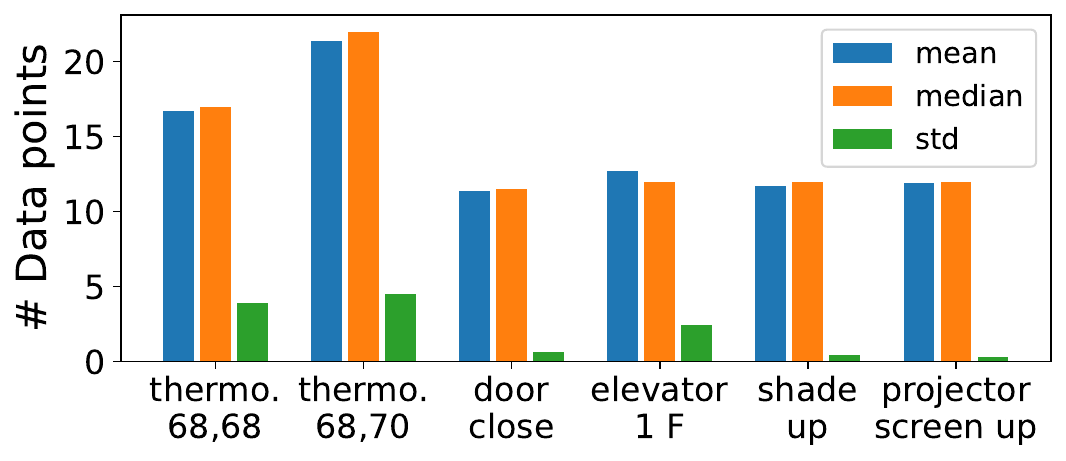}
        \vspace*{-0.3cm} 
        \caption{\small \sysname{} convergence time.}
        \label{fig:convergence}
        \vspace*{-0.3cm} 
    \end{subfigure}%
    \par\medskip
    \begin{subfigure}{\linewidth}
        \centering
        \includegraphics[width=0.8\linewidth]{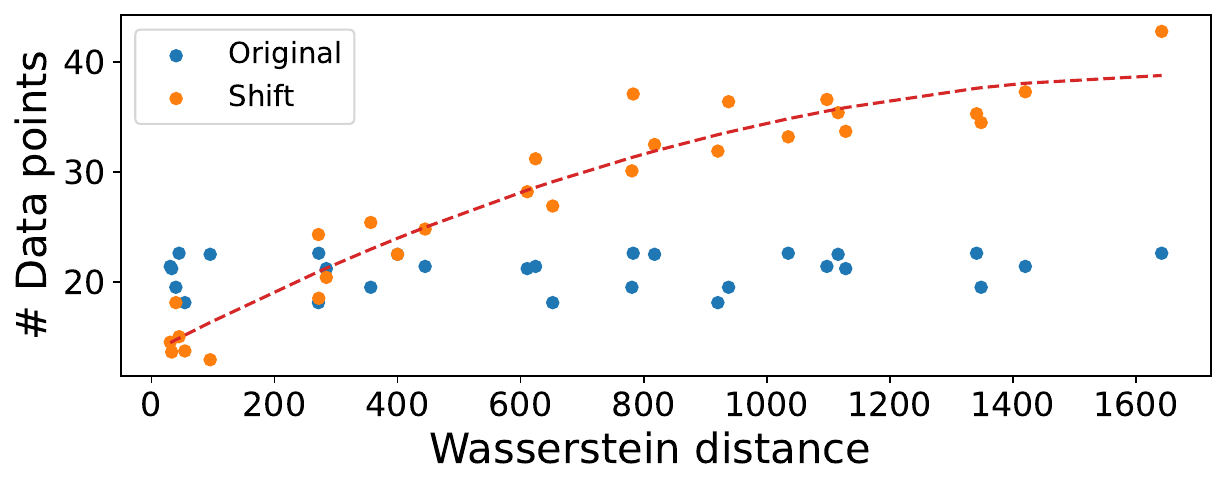}
        \vspace*{-0.3cm} 
        \caption{\small \sysname{} convergence  when action length drifts.}
        \label{fig:distribution-shift}
        \vspace*{-0.3cm} 
    \end{subfigure}
    \caption{\bf \small \sysname{} Convergence. \it \sysname{} converges quickly, requiring just over 20 (new) data points. }
    \label{fig:rasc-convergence}
    \vspace*{-0.5cm} 
\end{figure}

\begin{figure}[th!]
    \begin{subfigure}{\linewidth}
        \centering
        \includegraphics[width=0.7\linewidth]{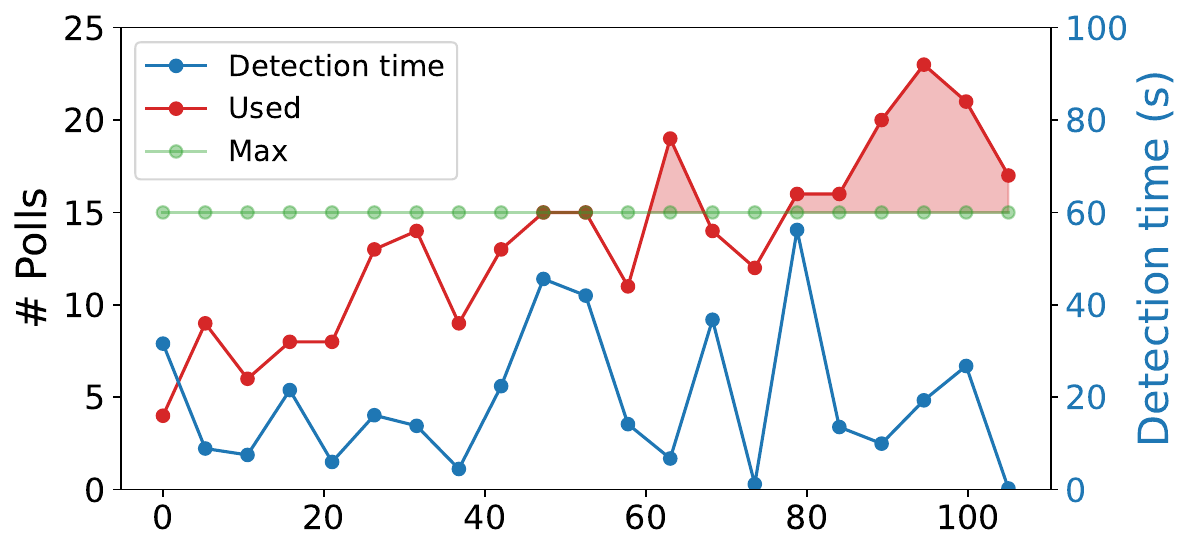}
        \vspace*{-0.3cm} 
        \caption{\small The interruption happens at the 50\% of the action.}
        \label{fig:action-interruption-fixed}
    \end{subfigure}%
    \vspace{-0.2cm}
    \par\medskip
    \begin{subfigure}{\linewidth}
        \centering
        \includegraphics[width=0.7\linewidth]{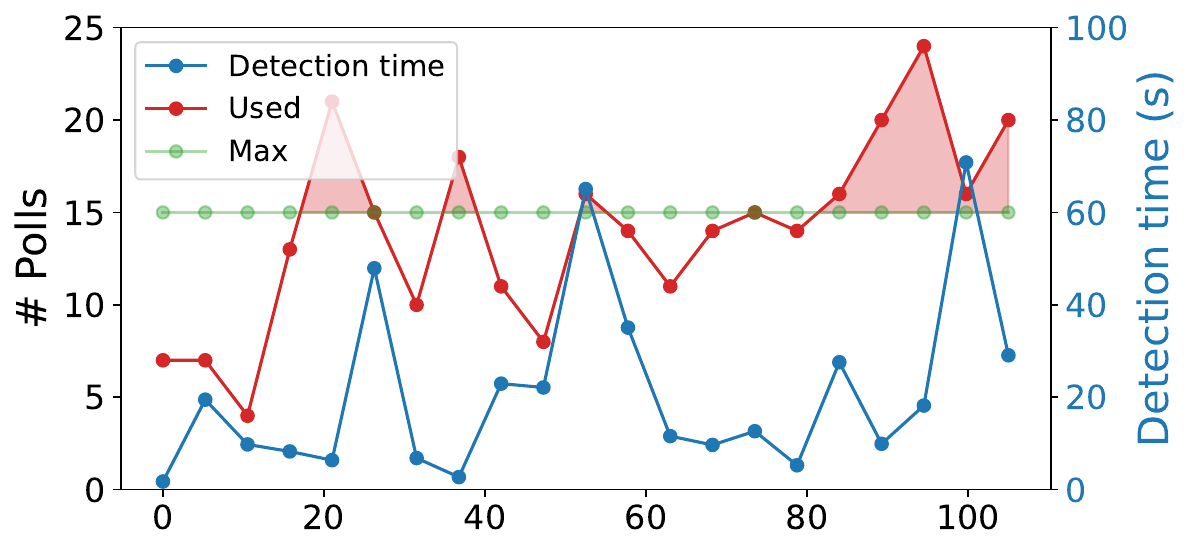}
        \vspace*{-0.3cm} 
        \caption{\small The interruption happens at the 80\% of the action.}
        \label{fig:action-interruption-fixed-80}
    \end{subfigure}
    \vspace*{-0.7cm} 
    \caption{\bf \small \sysname{} performance (polls) vs. Interruption Length relative to action length (\%). 
    Action: thermostat from 68 to 69. {\it \sysname{} generally uses fewer polls than max. Exceptions: when interruptions are long (right side of plots).}}
    \label{fig:action-interruption}
    \vspace*{-0.45cm} 
\end{figure}

\begin{figure*}[t!]
    \centering
    \includegraphics[width=\linewidth]{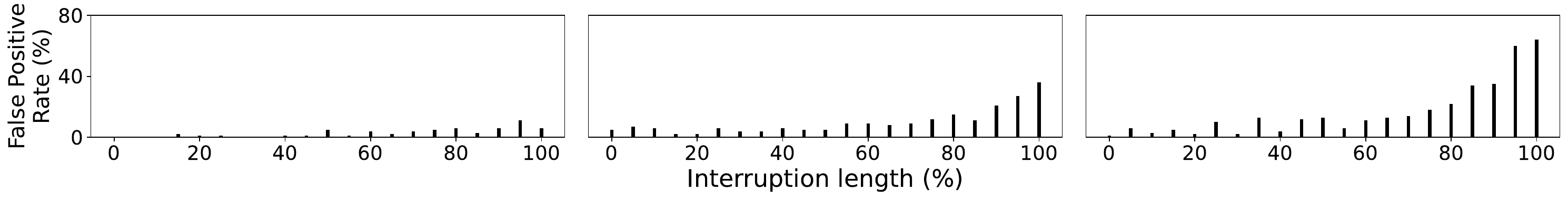}
    \vspace*{-0.8cm} 
    \caption{\bf \small Action Failure False Positive Rate during interruption. Left to right: interruption happens (resp.) at the 50\%, 80\%, and 90\% points of action. {\it False positives are generally low. Exception: when interruptions are long and towards the action's end. }}
    \label{fig:action-interruption-simulated}
    \vspace*{-0.5cm} 
\end{figure*}

\noindent \textbf{Data Drift.} 
As devices get older, a single action's probability distributions for its progress points may start moving towards different ranges and probabilities.
How quickly can \sysname{}'s measurements catch up?  \Figure~\ref{fig:distribution-shift} measures (for a thermostat device, at various temperature settings) the Wasserstein (earthmover) distance~\cite{vaserstein1969,kantorovich1942}. This metric measures the distance between the real and learned distributions, for both the original distribution and the shifted distribution. We observe quick convergence, especially when the original distribution is $< 500$ (already a large value, much higher than the error in our measurements)  
For instance, the Wasserstein distance between the distributions from the door and the elevator is approximately 13.93. 
In reality, it is highly unlikely that a door would experience such significant degradation.

\subsection{Detecting Interrupted Actions}
\label{sec:interruption-eval}

Interruptions to ongoing actions are common due to environmental factors and human interactions.
\Figure~\ref{fig:action-interruption} shows a real deployment where a thermostat change (68$\rightarrow$69) is interrupted at 50\% and 80\% progress. 
The red area denotes extra polls beyond $U$. 
When interruptions occur earlier or are brief, \sysname’s overhead remains low; with prolonged interruptions (right side of both plots), overhead rises, but detection time (blue line) stays low.

\Figure~\ref{fig:action-interruption-simulated} shows a simulation where we focus on {\it False Positive rate} (last row), i.e., the percentage of experiments where \sysname{} {\it falsely} detected an action failure.
We observe the False Positive rate rises with longer interruptions, especially when they occur near the action’s end, indicating \sysname{} reliably distinguishes true failures from interruptions in most cases.

\subsection{Routine Scheduling}
\label{sec:sched-finetuning}

\begin{figure}[t!]
    \vspace{-0.2cm}
    \begin{subfigure}{0.49\linewidth}
        \centering
        \includegraphics[width=\linewidth]{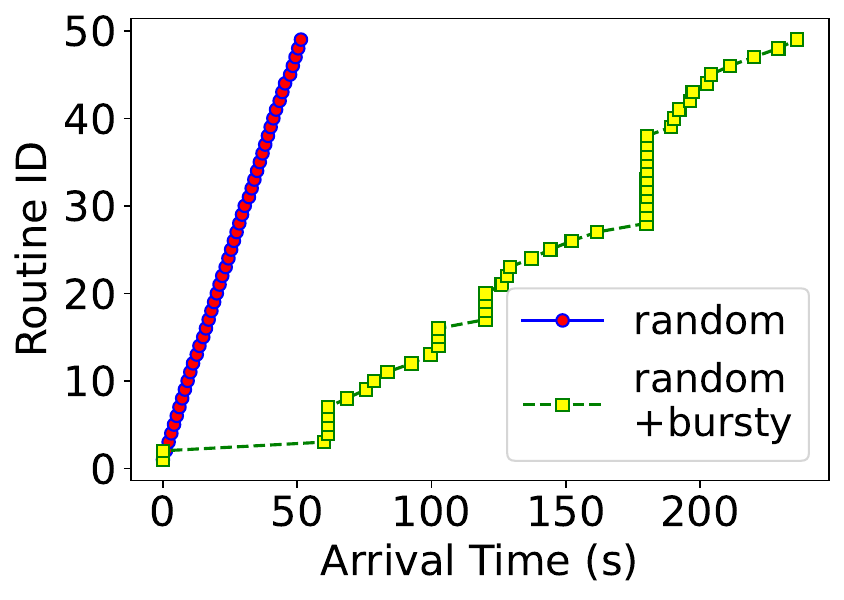}
        \captionsetup{width=.9\linewidth}
        \vspace*{-0.5cm} 
        \caption{\small The timeline of routine arrivals for the two datasets.}
        \label{fig:routine-arrival}
    \end{subfigure}%
    \hfill
    \begin{subfigure}{0.49\linewidth}
        \includegraphics[width=\linewidth]{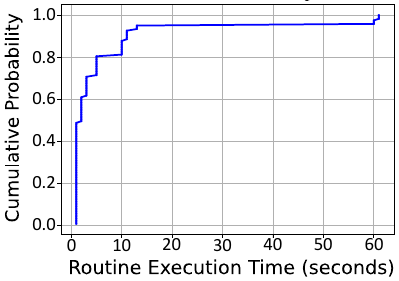}
        \caption{\small The sum of action lengths per routine as a CDF plot. }
        \label{fig:routine-action-lengths}
    \end{subfigure}
    \vspace{-0.4cm} 
    \caption{\bf \small Our Routine Benchmark. 
    }
    \label{fig:routine-dataset}
    \vspace{-0.3cm} 
\end{figure}

We evaluate \sysname's end-to-end performance: overheads and comparing \scheduler{} scheduling (\Section~\ref{sec:dyn-sched})
to baselines.

\noindent {\bf Dataset of Routines.}
We generated a suite of routines spanning diverse devices and lengths---statistics of our routine dataset are shown in \Figure~\ref{fig:routine-dataset}.
We have two
arrival workloads: (1) {\it random}, attempting to capture ``background'' routine arrival throughout the 24 hours of a day, 
and (2)  {\it random+bursty}, a mix of 50\% random and 50\% bursty arrivals, 
intended to capture ``peak'' times of activity, such as morning, lunch, evening.

\begin{table}[t]
    \centering
    \footnotesize
    \setlength{\tabcolsep}{5pt}
    \begin{tabular}{|c|c|c|c|c|c|c|c|}
        \hline
        \multirow{2}{*}{Dataset} & \multirow{2}{*}{\parbox{1cm}{\centering Polling Strategy}} & \multicolumn{3}{c|}{CPU (\%)} & \multicolumn{3}{c|}{Memory (MB)} \\
        \cline{3-8}  
         & & avg & q50 & q99 & avg & q50 & q99 \\
        \hline\hline
        \multirow{3}{*}{random} & adaptive & \textbf{1.96} & 1.6 & \textbf{5.53} & \textbf{0.25} & \textbf{0.25} & \textbf{0.5} \\
         & periodic & 2.95 & 2.0 & 15.09 & 0.4 & 0.38 & 0.88 \\
         & none & 2.52 & \textbf{1.2} & 22.86 & 25.49 & 25.75 & 26.0 \\
        \hline
        \multirow{3}{*}{\parbox{0.85cm}{\centering random + bursty}} & adaptive & 0.95 & 0.6 & 5.09 & \textbf{0.12} & \textbf{0.0} & \textbf{0.5} \\
         & periodic & 1.17 & 0.7 & 5.96 & 0.83 & 0.75 & 1.12 \\
         & none & \textbf{0.76} & \textbf{0.5} & \textbf{3.01} & 13.88 & 14.0 & 14.12 \\
        \hline
    \end{tabular}
    \vspace{-0.3cm}
    \caption{\bf \small \sysname's resource utilization under different polling strategies. Bold font indicates the lowest value for each dataset/metric combination. {\it \sysname{} either incurs the lowest CPU or memory (among baselines), or is only marginally higher than the lowest option.}}
    \label{tab:resource-overheads}
    \vspace{-0.6cm}
\end{table}

\noindent \textbf{Overheads.}
\Table~\ref{tab:resource-overheads} reports \sysname{}’s process overhead under different polling strategies. 
Adaptive polling is consistently more efficient than periodic---up to 63\% lower CPU and 43\% lower memory. 
Disabling polling (\emph{none}) yields only marginal or inconsistent gains (CPU q99 up to 68\%, otherwise up to 33\%), and memory can be worse. {In fact, memory is higher with no polling than with \sysname{}, indicating our automation-tracking memory management outperforms Home Assistant’s baseline, which overuses external libraries.}  
Thus, adaptive polling remains lightweight and effective.

\begin{figure*}[t]
  \centering

  \begin{minipage}[b]{0.39\textwidth}
    \centering
    \includegraphics[width=0.7\linewidth]{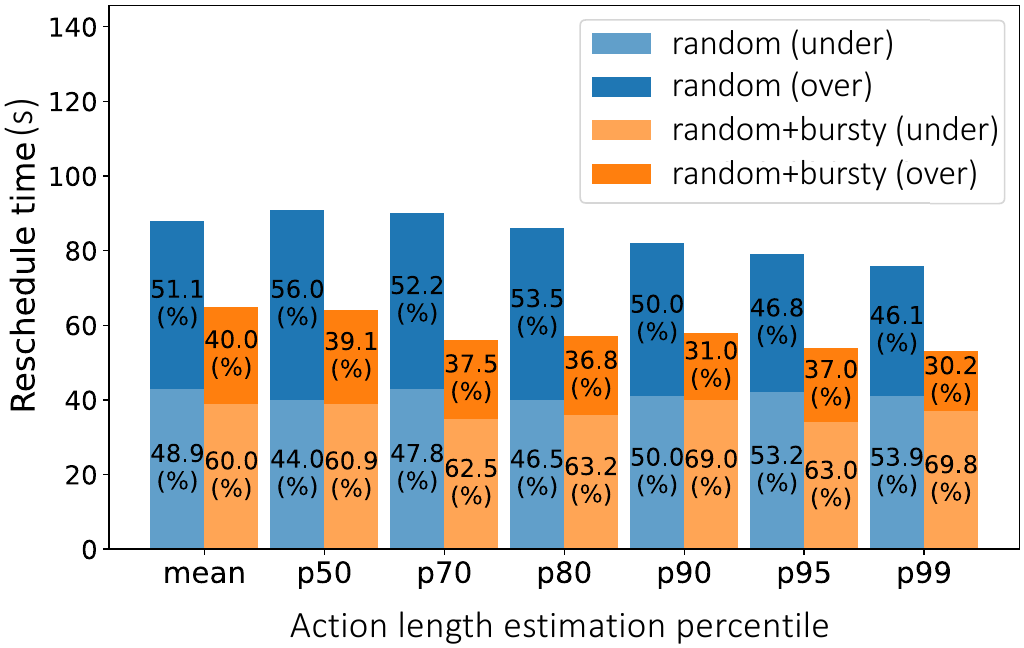}
    \vspace{-0.35cm}
    \captionof{figure}{\bf \small Rescheduling Overhead.
    Bars show total rescheduling time over the dataset, split into under-/over-time handling. {\it \sysname{} overhead stays low.}}
    \label{fig:reschedule_overhead}
  \end{minipage}
  \hfill
  \begin{minipage}[b]{0.24\textwidth}
    \centering
    \includegraphics[width=\linewidth]{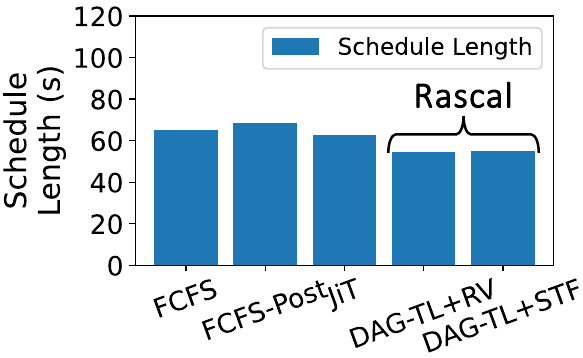}
    \vspace{-0.55cm}
    \captionof{figure}{\small \bf Schedule length comparison.
    {\it \sysname{} (right two) creates at least 11\% faster schedules vs. Baselines (left three).}
    }
    \label{fig:sched_len_comparison}
  \end{minipage}
  \hfill
  \begin{minipage}[b]{0.32\textwidth}
    \centering
    \includegraphics[width=\linewidth]{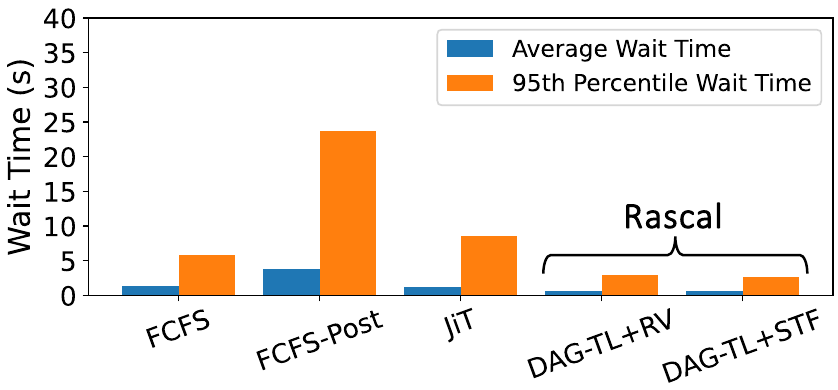}
    \vspace{-0.55cm}
    \captionof{figure}{\small \bf 
    Wait time comparison. {\it \sysname{} (right two groups) decreases wait times by at least 33\% and 50\% for mean and q95 values, respectively, vs. Baselines (three left groups).}
    }
    \label{fig:wait_time_comparison}
  \end{minipage}

  \vspace{-0.4cm}
\end{figure*}

\Figure~\ref{fig:reschedule_overhead} shows \sysname's rescheduling overheads with STF (Shortest Task First). For both datasets, rescheduling time drops 
as action-length estimates increase, indicating that overestimation reduces overhead---even with more early completions. 
The random+bursty dataset achieves faster rescheduling time as the schedule is tighter at times and traversals are faster. 

\label{sec:sched-comparisons}

\begin{figure*}[t]
    \begin{subfigure}{0.36\textwidth}
        \centering
        \includegraphics[width=0.9\linewidth]{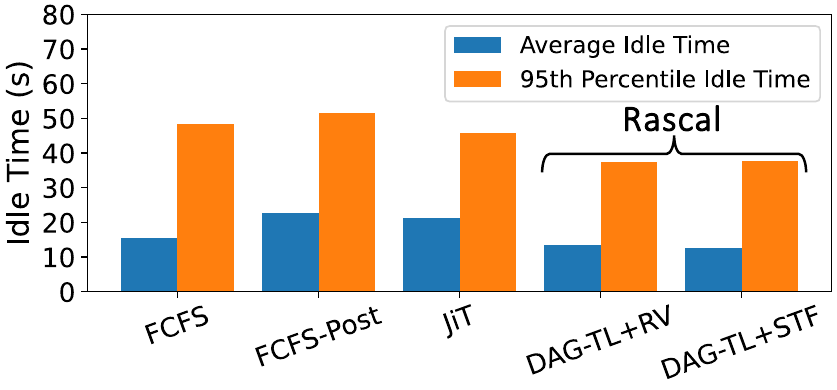}
        \vspace{-0.3cm} 
        \caption{\small \bf Idle time comparison. \it \sysname{} achieves at least 27\% and 18\% lower mean and q95 values, resp.}
        \label{fig:idle_time_comparison}
    \end{subfigure}
    \hfill
    \begin{subfigure}{0.36\textwidth}
        \centering
        \includegraphics[width=0.9\linewidth]{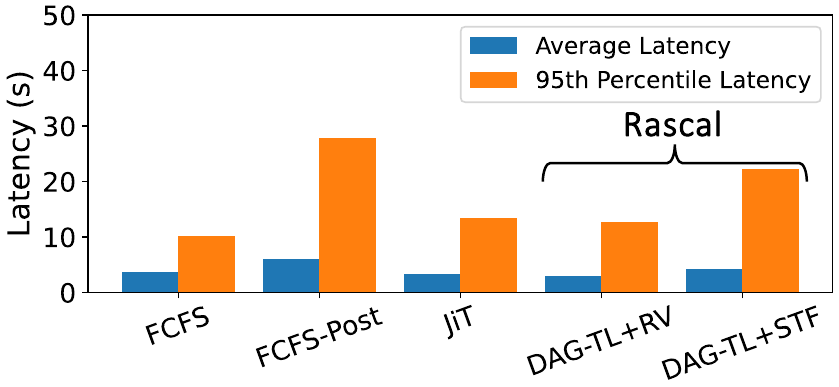}
        \vspace{-0.3cm} 
        \caption{\small \bf Latency comparison. \it \sysname{} achieves at least 25\% and 55\% lower mean and q95 values, resp.}
        \label{fig:latency_comparison}
    \end{subfigure}
    \hfill
    \begin{subfigure}{0.25\textwidth}
        \centering
        \includegraphics[width=\linewidth]{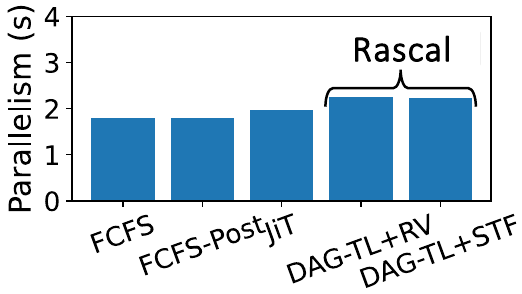}
        \vspace{-0.7cm} 
        \caption{\small \bf Parallelism comparison. \it \sysname{} achieves $>$ 10\% parallelism.}
        \label{fig:parallelism_comparison}
    \end{subfigure}
    \vspace*{-0.4cm} 
    \caption{\bf \small Routine Benchmark on \sysname{} (right two bars in each plot) vs. Baselines (left three bars).} 
    \vspace*{-0.5cm} 
    \label{routineresults1}
\end{figure*}

\noindent \textbf{\sysname{} Rescheduler Vs. Baselines.} We compare two variants of \sysname{} reschedulers (STF and RV from \Section~\ref{sec:rescheduler}: labeled in our plots as \scheduler+STF and \scheduler+RV respectively) against three baselines. 
Our three baselines are: (i)  FCFS (First Come First Served): earlier arriving routines have their actions scheduled earlier; (ii) FCFS-Post: a faster FCFS variant allowing a routine to start if the 
previous routine by FCFS standards has finished executing on all common devices; (iii) Just in Time (JiT) algorithm ~\cite{SafeHomeEuroSys2021}, which is a greedy and opportunistic approach.

Figs.~\ref{fig:sched_len_comparison}, \ref{fig:wait_time_comparison} show that 
(i) \sysname's two variants (\scheduler+RV and \scheduler+STF) perform similarly,
(ii) \sysname's slowest variant (\scheduler+STF) finishes routines 11\% faster than the fastest baseline (JiT), and
(iii) \sysname's slowest variant's wait times are also lower---mean by 33\% and q95 by 50\%---than the strongest baselines (JiT and FCFS, respectively). 
\Figure~\ref{fig:idle_time_comparison} shows that \sysname{} reduces idle time (time a device is not utilized) by 27\% at mean and 18\% at 95th percentile compared to the fastest baseline (FCFS and JiT, respectively).
In \Figure~\ref{fig:latency_comparison}, \sysname{}'s slowest variant increases the routine end-to-end latency average time by only 25\% and 95th percentile by 55\%, compared to the fastest baseline (JiT and FCFS, respectively).
Finally, \Figure~\ref{fig:parallelism_comparison} shows that \sysname{}'s least efficient variant achieves 10\% higher average parallelism rates over the most efficient baseline (JiT). 
Overall, \sysname's rescheduling improves metrics by 10–55\% over baselines.

\vspace{-0.3cm}

\section{Discussion} 

While all IoT devices we encountered show stable action time distributions, some are inherently variable---HVAC varies seasonally; oven and refrigerator times depend on load, etc. Handling such non-stationarity transparently remains challenging for IoT research; \sysname’s estimators are a step toward this goal. More expressive versions of \sysname{} programs 
can be explored, if their marginal benefit is high enough compared to the complexity they might add.

Another issue is the limited causality expressiveness in current automation interfaces (e.g., Alexa~\cite{Alexa}, Google Home~\cite{GoogleHome}).
These platforms center on IFTTT-style~\cite{ifttt} rules that struggle with conditional, multi-stage dependencies, limiting smart spaces---especially in larger, more dynamic environments. Current scripting lacks constructs for nuanced behaviors driven by multiple conditions or real-time updates, pushing users toward suboptimal automations. Future work should integrate more expressive, flow-based (visual) systems such as Node-RED~\cite{NodeRED} and Flogo~\cite{flogo}.

\section{Related Work}
\label{sec:related-work}

{
\noindent \textbf{Scheduling and Action Dependencies.}  
A range of systems provide abstractions for coordinating IoT and cyber-physical devices. 
Gaia~\cite{gaia}, Bundle~\cite{vicaire2010bundle}, dSpace~\cite{fu2021dspace}, and DepSys~\cite{depsys} offer middleware or dependency-based programming, but at coarse granularity without action-level progress or dynamic scheduling. 
TransActuations~\cite{transactuations}, Rivulet~\cite{rivulet}, and IoTRepair~\cite{iotrepair2020} strengthen safety through transactional or rollback semantics, yet operate only at the level of instantaneous or fixed-length actions or routines. 
SafeHome~\cite{SafeHomeEuroSys2021} and Hades~\cite{hades} handle routine conflicts but not fine-grained dependencies. 
In contrast, \sysname{} captures dependencies within as well as across routines, and our schedulers ensure safety and serial equivalence even as actions progress unpredictably.

\noindent \textbf{Dynamic Rescheduling.}  
Classical resource-reclaiming algorithms in real-time systems~\cite{shen1993resource,manimaran1997new,gupta2000new} reclaim slack when tasks finish early, but assume fixed DAGs and do not address late completions. 
Our adaptation of RV and STF extends these ideas to IoT action graphs with cross-routine serialization constraints, handling both early and delayed completions without violating safety.

\noindent \textbf{Observability and Failure Detection.}  
IoT systems often rely on coarse timeout-based detection or assume instantaneous actions~\cite{kashef2021wireless}, which is unsuitable for long-running tasks. 
SafeHome~\cite{SafeHomeEuroSys2021} and Hades~\cite{hades} provide some exception handling but not progress-aware detection. 
Our adaptive polling provides fine-grained observability of start/complete events, enabling efficient and accurate failure detection without overloading the system.
While pub/sub systems~\cite{gryphon,carzaniga2001design,aguilera1999matching,segall1997elvin} are often proposed as alternatives, they are best suited for settings with many subscribers across diverse topics. \sysname{} runs on a single central hub that subscribes to all topics, so pub/sub offers little benefit, and polling remains the core challenge it addresses.
The histogram bucket problem~\cite{scott1979optimal,freedman1981histogram,knuth2006optimal,scargle2013studies,jagadish1998optimal} is analogous: both tasks require selecting partition ranges (bins vs. polling intervals) under uncertainty, differing only in the optimization criterion—e.g., minimizing error in density estimation versus balancing observability cost and timeliness.

\noindent \textbf{Commercial Systems.}  
Commercial platforms such as Alexa~\cite{Alexa}, Google Home~\cite{GoogleHome}, SmartThings~\cite{SmartThings}, and Home Assistant~\cite{HomeAssistant} support limited forms of automation and do not guarantee conflict freedom or robust failure handling. 
\abs{} generalizes them for expressive DAG-based routines, with progress awareness and principled scheduling guarantees.
}

\section{Conclusion}

\abs{} provides an expressive RPC alternative for IoT device collections such as smart homes, smart buildings, etc. Our implementation, \sysname, supports \textit{observability} by detecting action completion and failures quickly, and \textit{programmability} via fine-grained causality dependencies, and assures safety and serial-equivalence. \sysname{} requires no modifications to existing devices. 
Trace-driven evaluation showed that \sysname{} detects completion within 2-13 RPCs and 2-16s over 90\% of the time,
even for actions lasting tens of minutes.  Our
routine scheduling outperforms state-of-the-art baselines by 10-55\%.

\section*{Acknowledgments}
This work was supported in part by NSF grant CNS 1908888, NSF grant CNS 2504595, an IIDAI grant, and gifts from Microsoft and Capital One.

This paper is a preprint version of our upcoming paper of the same name in the USENIX Symposium on Networked Systems Design and Implementation (NSDI), 2026.

\bibliographystyle{plain}
\bibliography{12_bibliography}

\appendix

\section{Extended Analysis of the Adaptive Polling Algorithm}
\label{app:rasc-proofs}

\begin{reptheorem}[Adaptive Poll Placement with Fixed Budget $k$]{1}
Given a time distribution $p(t)$ on $(0,U]$, a poll budget $k$, and a terminal tolerance $\varepsilon>0$,
polls
$0<{L_1^{\!*}}<\cdots<L_{k-1}^{\!*}<L_k^{\!*}$
that minimize
expected detection time $Q$ satisfy $\lvert L_k^{\!*}-U\rvert\le\varepsilon$ are given by the following recurrent relation:
\[
L_i^{\!*} =
\begin{cases} 
\frac{1}{p(L_{i-1}^{\!*})} \cdot \int_{L_{i-2}^{\!*}}^{L_{i-1}^{\!*}} p(t)\, dt + L_{i-1}^{\!*} & \text{for } i \in \{2,\dots,k\}, \\
\text{value } \in (0,U] & \text{for } i=1
\end{cases} \tag{5} \label{eq:rep-recurrence}
\]
\end{reptheorem}

\begin{proof}
Only the $i$-th and $(i+1)$-th terms of $Q$ depend on $L_i$:
\[
A_i=\int_{L_{i-1}}^{L_i} (L_i-t)\,p(t)\,dt,\qquad
A_{i+1}=\int_{L_i}^{L_{i+1}} (L_{i+1}-t)\,p(t)\,dt.
\]
By Leibniz’s rule,
\[
\frac{\partial A_i}{\partial L_i}
= (L_i{-}L_i)p(L_i) + \int_{L_{i-1}}^{L_i} \frac{\partial}{\partial L_i}\big[(L_i{-}t)p(t)\big]\,dt
= \int_{L_{i-1}}^{L_i} p(t)\,dt,
\]
and
\[
\frac{\partial A_{i+1}}{\partial L_i}
= -\,(L_{i+1}-L_i)\,p(L_i).
\]
The first-order optimality condition $\partial Q/\partial L_i=0$ gives
\[
\int_{L_{i-1}}^{L_i} p(t)\,dt \;-\; (L_{i+1}-L_i)\,p(L_i) \;=\;0,
\]
which rearranges to \eqref{eq:rep-recurrence}.
Under $p>0$ and continuity, these conditions determine a unique sequence once $L_1$ is fixed.

Fix $L_1\in(0,U)$ and generate $L_2,\dots,L_k$ via \eqref{eq:rep-recurrence}.
Then $L_k(L_1)$ is strictly increasing in $L_1$.
Hence, there exists a unique $L_1^*\in(0,U)$ such that the generated sequence satisfies $L_k^*=U$.
\end{proof}

\begin{reptheorem}[Meeting a Detection Tolerance SLO]{2}
Given a detection window $Q_w\in(0,U]$, an SLO $\text{slo}\in(0,1]$, and
a placement $\mathcal{L}=\{L_1<\dots<L_k=U\}$, its covered set is
\[
C(\mathcal{L})=\bigcup_{i=1}^{k}\big((L_i-Q_w)^+,\,L_i\big]\cap(0,U],
\quad (x)^+=\max\{x,0\},
\]
and its coverage is
\[
\mathrm{Cover}(\mathcal{L})=\int_{C(\mathcal{L})}\!p(t)\,dt.
\]
Let $A_k(Q_w)$ denote the best achievable coverage with $k$ polls,
\[
\begin{aligned}
A_k(Q_w) &= \max_{\mathcal{L}}\, \mathrm{Cover}(\mathcal{L}),\\
k^*(\text{slo},Q_w) &= \min\{\,k\in\mathbb{N}: A_k(Q_w)\ge \text{slo}\,\}.
\end{aligned}
\]

Assume a binary search over $k$ is run with initial bracket $\texttt{left}=0$, $\texttt{right}=\lceil U/Q_w\rceil$,
using an oracle \texttt{examineQw} that returns \texttt{true} exactly when there exists a placement $\mathcal{L}$ with $\mathrm{Cover}(\mathcal{L})\ge \text{slo}$.
Then the search returns $k^*(\text{slo},Q_w)$ together with an SLO-feasible placement.
\end{reptheorem}

\begin{proof}
\noindent\textbf{Monotonicity.} If $k'<k$, any $k'$-poll placement can be extended to $k$ polls by inserting additional polls; this cannot reduce $C(\mathcal{L})$, hence $A_k(Q_w)\ge A_{k'}(Q_w)$. The oracle predicate is therefore nondecreasing in $k$.

\noindent\textbf{Bracketing.} For $k=\lceil U/Q_w\rceil$, equal spacing $L_i=iU/k$ yields $\bigcup_i ((L_i-Q_w)^+,L_i]\supseteq(0,U]$, so $\mathrm{Cover}(\mathcal{L})=1\ge \text{slo}$. Thus the upper bracket is feasible, while $k=0$ is not.

\noindent\textbf{Minimality via binary search.} Binary search on a nondecreasing predicate with a feasible upper bracket returns the smallest $k$ for which the predicate holds; by definition this is $k^*(\text{slo},Q_w)$.
\end{proof}

\balance

\end{document}